%% file: HEFT_hilbert.tex
\documentclass[a4paper,11pt]{article}
\pdfoutput=1

\usepackage{jheppub} 
\usepackage[T1]{fontenc} 
\usepackage{slashed}
\usepackage{braket}
\usepackage{color}
\usepackage{amsmath,amssymb,amsfonts,booktabs}
\usepackage{tikz}
\usetikzlibrary{decorations.pathreplacing}

\usepackage{subfigure}
\usepackage{multirow}
\usepackage{array}
\usepackage{arydshln}
\usepackage{ytableau}

\usepackage{graphicx}
\usepackage{cleveref}
\usepackage{hyperref}
\usepackage{srcltx}
\usepackage{bbm}
\usepackage{mathdots}
\usepackage{dsfont}

\crefname{table}{Table}{Tables}
\crefname{equation}{Eq.}{Eqs.}
\crefname{appendix}{App.}{Apps.}
\crefname{section}{Sec.}{Secs.}
\crefname{figure}{Fig.}{Figs.}

\input{preamble.tex}

\newcommand\blue[1]{\textcolor{blue}{#1}}  

\def\cD{{\cal D}}

\def\cO{{\cal O}}

\def\Hcal{\mathcal{H}}
\renewcommand{\a}{\alpha}
\renewcommand{\b}{\beta}

\newcommand{\hc}{\text{h.c.}}

\newcommand{\Qbar}{{\bar Q}}
\newcommand{\Lbar}{{\bar L}}
\def\beq{\begin{equation}}
\def\eeq{\end{equation}}

\newcounter{twofermionQ}

\newcounter{twofermionL}

\newcounter{fourfermionQ}

\newcounter{fourfermionL}

\newcounter{fourfermionQL}

\newcommand{\ie}{\textit{i.e.}}
\newcommand{\eg}{\textit{e.g.}}

\def\tcb#1{\textcolor{blue}{#1}}
\def\tcm#1{\textcolor{magenta}{#1}}

\title{\boldmath Hilbert Series, the Higgs Mechanism, and HEFT}

\author[a,b]{Luk\'a\v{s} Gr\'af,}\emailAdd{lukas.graf@berkeley.edu}
\author[c,d]{Brian Henning,}\emailAdd{brian.henning@epfl.ch}
\author[b,e]{Xiaochuan Lu,}\emailAdd{xil224@ucsd.edu}
\author[f]{Tom Melia,}\emailAdd{tom.melia@ipmu.jp}
\author[a,f,g,1]{Hitoshi Murayama\note{Hamamatsu Professor}}\emailAdd{hitoshi@berkeley.edu}

\affiliation[a]{Department of Physics, University of California, Berkeley, CA 94720, USA}
\affiliation[b]{Department of Physics, University of California, San Diego, La Jolla, CA 92093, USA}
\affiliation[c]{Institute of Physics, Theoretical Particle Physics Laboratory (LPTP), \'Ecole Polytechnique F\'ed\'erale de Lausanne (EPFL), CH-1015 Lausanne, Switzerland}
\affiliation[d]{D\'epartment de Physique Th\'eorique, Universit\'e de Gen\`{e}ve, CH-1211 Gen\`eve, Switzerland}
\affiliation[e]{Institute for Fundamental Science, Department of Physics, University of Oregon, Eugene, OR 97403, USA}
\affiliation[f]{Kavli Institute for the Physics and Mathematics of the
  Universe (WPI), University of Tokyo Institutes for Advanced Study, University of Tokyo,
  Kashiwa 277-8583, Japan}
\affiliation[g]{Ernest Orlando Lawrence Berkeley National Laboratory, University of California, Berkeley, CA 94720, USA}

\abstract{We expand Hilbert series technologies in effective field theory for the inclusion of massive particles, enabling, among other things, the enumeration of operator bases for non-linearly realized gauge theories. We find that the Higgs mechanism is manifest at the level of the Hilbert series, as expected for the  partition function of an $S$-matrix that is subject to the Goldstone equivalence theorem.  In addition to massive vectors, we detail how other massive, spinning particles can be studied with Hilbert series; in particular, we spell out the ingredients for massive gravity in general spacetime dimensions. Further methodology is introduced  to enable Hilbert series to capture the effect of spurion fields acquiring vevs. We apply the techniques to the Higgs Effective Field Theory (HEFT), providing a systematic enumeration of its operator basis. This is achieved both from a direct and a custodial symmetry spurion-based approach; we compare and contrast the two approaches, and our results to those appearing in previous literature.}

\begin{document} 
\maketitle
\flushbottom
\setcounter{page}{2}

\section{Introduction}
\label{sec:intro}

Partition functions are fundamental objects of study in physics as they provide information about the energy spectrum of a theory and its degeneracies. Similarly, they can be constructed to provide the spectrum of allowed scattering elements and degeneracies of the $S$-matrix of a quantum Effective Field Theory (EFT). Such partition functions can be identified as Hilbert series~\cite{Henning:2017fpj}. Because operators are related to scattering observables,  with each operator corresponding to a contact scattering interaction, Hilbert series can equivalently be seen as solving the operator construction problem for an EFT. The methodology has been developed to cover a broad class of phenomenological EFTs~\cite{Jenkins:2009dy,Hanany:2010vu,Henning:2015daa,Lehman:2015via,Lehman:2015coa,Henning:2015alf,Henning:2017fpj,Kobach:2017xkw,Kobach:2018nmt,Kobach:2018pie,Henning:2019mcv,Henning:2019enq,Ruhdorfer:2019qmk, Graf:2020yxt}. Most recently, a Hilbert series was constructed for the pion/chiral Lagrangian of QCD~\cite{Graf:2020yxt}, the archetypical example of an EFT with a non-linearly realized, spontaneously broken symmetry.

The power of chiral Lagrangians lies in their exploitation of symmetry and the Nambu-Goldstone theorem: the low energy dynamics is described by interactions of massless bosons. The famous exception is when the symmetry is a gauge symmetry and  gauge bosons acquire mass via the Higgs mechanism.  Hilbert series for this important class of phenomenological theory have not yet been studied. The purpose of the current paper is to fill this gap and detail their construction. We will  use  the Higgs Effective Field Theory (HEFT, or its extension with right handed neutrinos, $\nu$HEFT) that describes electroweak symmetry breaking in the standard model as an example to which we apply the techniques and  compare with previous literature.

The broad idea of building an effective field theory is to first identify the relevant degrees of freedom---degrees of freedom which are directly accessible, \textit{i.e.} on-shell---and then constrain the allowed interactions using principles the system adheres to, such as locality and various possible symmetries. In many cases the relevant degrees of freedom are massless, or nearly so, and are low-energy manifestations of heavier dynamics that has been integrated out. However, EFTs can be equally useful to describe physics within an energy window: the upper edge of the window signifies the breakdown of the EFT (typically through some form of unitarity violation) and the need to include new degrees of freedom, while the lower edge suggests---but does not mandate---integrating out now inaccessible degrees of freedom and working with a new effective theory. In this regard, it is important to determine how massive particles are systematically incorporated into EFTs.

An important realization of the above scenario is provided by the electroweak sector of the Standard Model: massive electroweak gauge bosons are copiously produced at the LHC, but only leave virtual impacts in the beta decay processes studied at \href{https://www.katrin.kit.edu/}{KATRIN}. So while four-Fermi theory---or it's modern incarnation LEFT~\cite{Jenkins:2017jig}---is sufficient to study physics in Karlsruhe, physics in Geneva requires an effective description incorporating the massive gauge bosons, the so-called electroweak chiral Lagrangian (e.g.~\cite{Feruglio:1992wf}), or its modern update HEFT that includes the singlet, scalar Higgs field \(h\).\footnote{The Standard Model EFT (SMEFT) provides another valid description of electroweak gauge bosons. In this formulation electroweak symmetry is linearly realized with a Higgs doublet \(H\) which subsequently gets a vacuum expectation value and breaks the electroweak symmetry. Therefore, formally the EFT is built with \textit{massless} vectors which then become massive due to the now specified dynamics of \(H\). Instead, in this work our focus is on building EFTs where the vectors are massive from the get-go, as a result of some unspecified dynamics. See Refs. \cite{Alonso:2015fsp,Alonso:2016oah,Cohen:2020xca} for the difference between HEFT and SMEFT from a geometric point of view.}

In this work we study how to systematically incorporate massive particles into EFTs, specifically in regard to enumerating and constructing the operator basis at arbitrary order. Our focus in the main text will be on the inclusion of massive vectors---showcasing the method through a detailed treatment of HEFT---while in the appendix we extend the Hilbert series methods to incorporate massive particles of arbitrary spin.

As we will see, the basic prescription of ``specify the particles and their symmetries'' is precisely how the Hilbert series calculation proceeds: in particular, we need only to know that the \(W\) and \(Z\) gauge bosons furnish a massive representation of Poincar\'e symmetry together with their respective charges under electromagnetism \(U(1)_{\text{EM}}\).\footnote{This is directly analogous to Hilbert series calculations for Nambu-Goldstone bosons~\cite{Henning:2017fpj,Graf:2020yxt} describing the spontaneous breaking of a global symmetry \(G\) to a subgroup \(H\): once we have identified the appropriate field variable (the Cartan-Maurer form built from \(e^{i\pi/f}\)) we need only to impose invariance under the unbroken group \(H\).} The structure of massive particle representations are interesting in their own right, as they reflect well known phenomena such as the Higgs mechanism. We will see this reflected in the single particle partition function (the key ingredient entering the Hilbert series), where for the case of a spin-1 boson we have
\begin{equation}
Z_{\text{Massive vector}} = Z_{\text{Massless vector}} + Z_{\text{N-G Boson}} \,.
\label{eq:intromv}
\end{equation}

The existence of such a relation is guaranteed by the Goldstone equivalence theorem~\cite{Lee:1977yc,Lee:1977eg} that dictates how the high energy $S$-matrix has to behave. Namely, the longitudinal and transverse components of scattering massive vector bosons decouple, and the  amplitudes approach those describing the scattering of massless vectors and N-G bosons. Its appearance  emphasizes the physical significance of the Hilbert series as enumerating the degeneracies of the $S$-matrix. See also the construction of operator bases using massive on-shell techniques that has been explored in~\cite{Durieux:2019eor,Durieux:2020gip,Balkin:2021dko}. The presentation in this paper encodes the all-order structure about such constructions, and Goldstone equivalence. 

As in previous incarnations of the Hilbert series in EFT applications, the identification of the underlying representation theory affords generalizations. In particular, we detail how the Higgs mechanism works at the level of the $S$-matrix in general spacetime dimensions $d$, and with particles of general spin $k$.
This can be seen as a scintilla of a general $d$ and $k$ Goldstone equivalence theorem. 
The general expression is again 
found by appealing to the mode decomposition of massive and massless spinning particles in $SO(d)$, echoing the analysis of conformal-helicity duality found in~\cite{Henning:2019mcv,Henning:2019enq}. Schematically one obtains
\begin{equation}
Z_{\text{Massive spin }k} = \sum_{l=0}^k Z_{\text{Massless spin }l}^{\partial^{k-l}}\,\,\,\,  \,,
\end{equation}
where the superscript $\partial^m$ indicates the particle couples through $m$-derivative interactions (c.f.\, \cref{eq:MassiveSpinkSPMdecomp}). It would be interesting to develop these results in line with the construction and study of EFTs for massive gravity in higher dimensions (see~\cite{Ruhdorfer:2019qmk} for a Hilbert series treatment of massless gravity).

As for other generalizations: while we do not pursue it here, techniques to implement parity~\cite{Henning:2017fpj} and charge conjugation~\cite{Graf:2020yxt} can be readily implemented.  We emphasize that the full formalism is readily applied as a systematic way of studying operator bases for theories with any symmetry breaking pattern and particle content.

The application of Hilbert series techniques to the HEFT requires some additional machinery to make contact with the literature. It involves developing some tricks that utilize `helper' spurions in Hilbert series, since it is common practice in the community to 1) write HEFT operators in the form of custodial symmetry invariants, with spurions that get a vev to break back down to electromagnetism (EM),\footnote{As we review in~\cref{sec:hefths} (see~\cref{subsec:HEFT_hs_spurion} in particular), most treatments in the literature embed the electroweak symmetry breaking structure \(SU(2)_L\times U(1)_Y \to U(1)_{\text{EM}}\) inside the custodially symmetric symmetry breaking pattern \(SU(2)_L \times SU(2)_R \to SU(2)_V\) and use spurions to treat the explicit breaking of the custodial \(SU(2)_V\) down to its \(U(1)_V\) subgroup. As elaborated below and in~\cref{subsec:HEFT_hs_spurion}, this \(U(1)_V\) is \textit{not} \(U(1)_{\text{EM}}\), but rather a linear combination of \(U(1)_{\text{EM}}\) and \(U(1)_{B-L}\), and therefore this approach has inherent limitations and technically is not equivalent to the most general form of HEFT. For a treatment based purely on the coset \(SU(2)_L\times U(1)_Y /U(1)_{\text{EM}}\), see~\cite{Chanowitz:1987vj}.} and 2) treat two operators related by a  polynomial factor in the singlet Higgs field $h$ as equivalent. Doing so touches on deeper properties of Hilbert series, their grading, and the representation and invariant theory, and we explore and elaborate on these issues in generality.

Regarding the comparison to the literature, we have checked our results against three different groups, including \cite{Buchalla:2013rka,Krause:2016uhw}, \cite{Brivio:2016fzo},  and \cite{Sun:2022ssa}. We find agreement with the original listing of all NLO operators in~\cite{Buchalla:2013rka,Krause:2016uhw}, except for the four-fermion operators. The four-fermion ones are however less of interest here in the sense that they are independent of the EW Nambu-Goldstone sector, and are identical to those operators appearing in the Standard Model EFT (SMEFT) or Low-energy EFT (LEFT). In particular, we find agreement with the original listing of four fermion LEFT operators~\cite{Jenkins:2017jig}. 

The approaches of \cite{Brivio:2016fzo} and \cite{Sun:2022ssa} are to embed the EW symmetry breaking structure inside the custodially symmetric breaking structure, and the operators are constructed via the introduction of a custodial symmetry triplet spurion. This spurion then breaks the $SU(2)_V$ custodial symmetry to its $U(1)_V$ subgroup generated by $t^3_V$.  This $t^3_V$ is not the same as EM charge $Q$; preserving $t^3_V$ is only equivalent to preserving $Q$ for the $B-L=0$ sector. Comparing within the $B-L=0$ sector, we find disagreement with the quoted total  number of four fermion operators in the HEFT summary tables of~\cite{Sun:2022ssa} (we find agreement with the corresponding number for $\nu$HEFT).\footnote{As we were finishing writing this paper, we noticed that the HEFT NNLO operator classes were worked out by the same group of authors \cite{Sun:2022snw}. It would be interesting to make a detailed comparison, but it is beyond the scope of the current paper.}

This is the story of Hilbert series, Higgs, and HEFT: \cref{sec:Recap} provides a lightening recap of Hilbert series applied to EFT; \cref{sec:SPMs} explains how to incorporate massive vector bosons and elucidates how the Higgs mechanism appears in the Hilbert series; \cref{sec:Spurion} outlines the new spurion techniques we introduce in the Hilbert series to mod out singlet polynomial form factors and/or implement the effect of spurion fields getting vevs; \cref{sec:hefths} details the application to HEFT, along with summary tables; and, \cref{sec:Conclusions} gives an outlook on possible future developments of the ideas presented here and elsewhere. An appendix collects a number of mathematical results on the mode decomposition and characters for massive and massless spin $k$ particles in $d$ spacetime dimensions, and the manifestation of the Higgs mechanism at the level of character theory. Hilbert series for ($\nu$)HEFT are included as an ancillary file accompanying the paper.

\section{Hilbert Series Recap}
\label{sec:Recap}

In this section, we provide a recap of how to compute the Hilbert series that encodes the essential information of the operator basis. This also serves as an introduction of the basic language.

Although the Hilbert series has a definition that makes it applicable to a general class of mathematical problems, when we apply it to the case of presenting an operator basis, its practical evaluation can be achieved following a relatively simple procedure---one first considers all possible ways of multiplying all fields' components (including their derivatives), and then selects out combinations of these products that respect the given set of symmetries of the theory, \ie\ combinations that are singlet representations of the symmetry groups. Concretely, this selection can be elegantly achieved by making use of character orthogonality in group representation theory:
\begin{equation}
\Hcal_0 (\Phi, q) = \int\text{d}\mu_\text{Internal}^{}(y) \int\text{d}\mu_\text{Spacetime}^{}(x)\, \frac{1}{P(q, x)}\, Z(\Phi, q, x, y) \,.
\label{eqn:H0}
\end{equation}
A few remarks about this master formula are in below.
\begin{itemize}
\item Consider the linear space $R_\text{All}$ of all operators (\ie\ all possible ways of multiplying all the fields' components). It forms a representation under the symmetry groups. The quantity $Z(\Phi, q, x, y)$ in \cref{eqn:H0} is the (graded) character of this representation.\footnote{The character is given by the trace over the representation matrix.} Specifically in  the above, spacetime symmetry group elements are parameterized by variables collectively labeled $x$, and internal symmetry group elements $y$. The character is then a function of $(x, y)$. We further introduce the grading variables $q$ and $\Phi$ to keep track of each operator's mass dimension and field content. So this graded character $Z(\Phi, q, x, y)$ encodes the information about the representation of all possible operators.
\item By character orthogonality, the integral over the Haar measure of  the (compactified) Lorentz group together with the inverse of the ``momentum generating function'' $P(q,x)$ that accounts for translation invariance (see \cref{eqn:Pqx} for the detailed expression)
\[
\int\text{d}\mu_\text{Spacetime}^{}(x)\, \frac{1}{P(q,x)}
\]
selects out operators accounting for Poincar\'e symmetry. 
\item The integral over the internal symmetry Haar measure
\[
\int \text{d}\mu_\text{Internal}^{}(y)
\]
further selects out operators that are singlets under the given set of internal symmetries.
\end{itemize}
We refer the reader to Section 3 in \cite{Graf:2020yxt} for a detailed explanation of the various quantities in \cref{eqn:H0}, and \cite{Henning:2017fpj} for a comprehensive discussion on Hilbert series method. Here, let us just mention that the representation of all operators $R_\text{All}$ is clearly a direct product of the representations made out of each single field, so the character is given by their multiplication:
\begin{equation}
R_\text{All} = \bigotimes_i \Big[ \bigoplus_n\, \text{(anti-)sym}^n \left( R_{\Phi_i} \right) \Big]
\quad\implies\quad
Z = \prod_i \text{PE}_{(f)} \left( \Phi_i\, \chi_{\Phi_i}^{} \right) \,.
\label{eqn:Zdecompose}
\end{equation}
In the above PE stands for Plethystic Exponential. The (anti-) and (f) apply for the case of fermionic fields.
We call each building block $R_{\Phi_i}$ a Single Particle Module (SPM). Therefore, the key to working out the quantity $Z(\Phi, q, x, y)$ is to work out the character of each SPM:
\begin{equation}
\chi_{\Phi_i}^{} (q, x, y) = \chi_{\Phi_i}^\text{Spacetime}(q,x)\, \chi_{\Phi_i}^\text{Internal}(y) \,.
\end{equation}
The relevant details of SPMs $R_{\Phi_i}$ and their characters $\chi_{\Phi_i}$ for HEFT and similar types of theories will be discussed in \cref{sec:SPMs}.

\begin{table}[t]
\renewcommand{\arraystretch}{1.3}
\setlength{\arrayrulewidth}{.2mm}
\setlength{\tabcolsep}{1em}
\centering
\begin{tabular}{cccc}
\toprule
$\cO^{[\mu_1\cdots\mu_k]}$              & rank $k$ & dim $\Delta_\cO$ & $\Delta\Hcal$ \\
\midrule
$\varepsilon^{\mu\nu\rho\sigma}$        & 4  & 0 & $-q^4$                      \\
$\pd^\mu \phi_i$                        & 1  & 2 & $q^3\,\phi_i$               \\
$\phi_{[i} \pd^\mu \phi_{j]}$           & 1  & 3 & $q^4\,\phi_i\,\phi_j$       \\
$\chi_i^\dagger \bar\sigma^\mu \chi_j$  & 1  & 3 & $q^4\,\chi_i^\dagger\,\chi_j$ \\
$F_i^{\mu\nu}\;,\;\tilde{F}_i^{\mu\nu}$ & 2  & 2 & $-q^4\,(F_{iL}+F_{iR})$     \\
\midrule
$u_i^\mu$                                                      & 1 & 1 & $q^2\,u_i$           \\
$\varepsilon^{\mu\nu\rho\sigma} u_{i\sigma}$                   & 3 & 1 & $q^4\,u_i$         \\
$\varepsilon^{\mu\nu\rho\sigma} u_{i\rho} u_{j\sigma}$         & 2 & 2 & $-q^4\,u_i\,u_j$   \\
$\varepsilon^{\mu\nu\rho\sigma} u_{i\nu} u_{j\rho} u_{k\sigma}$& 1 & 3 & $q^4\,u_i\,u_j\,u_k$ \\
$u_{i\nu} F_j^{\mu\nu}\;,\; u_{i\nu} \tilde{F}_j^{\mu\nu}$     & 1 & 3 & $q^4\,u_i\,(F_{jL}+F_{jR})$ \\
\bottomrule
\end{tabular}
\caption{Generic co-closed but not co-exact form operators $\cO^{[\mu_1\cdots\mu_k]}$ that could be built out of scalars $\phi_i$, left-handed Weyl fermions $\chi_i$, field strengths $F_i^{\mu\nu}$, and linearized Goldstone fields $u_i^\mu$. We give the corresponding contributions to $\Delta\Hcal$ in the case of considering no internal symmetries. When there are internal symmetries on top of it, one follows \cref{eqn:DeltaH} to further select out internal symmetry singlets. The upper section collects form operators that do not involve Goldstone fields, whose contributions to $\Delta\Hcal$ can be actually worked out automatically using conformal representation theory as explained in \cite{Henning:2017fpj}. The lower section collects our manual enumeration of co-closed but not co-exact form operators that involve Goldstone fields.}
\label{tbl:CohomologyForms}
\end{table}

\cref{eqn:H0} is an elegant formula for carrying out the simple spirit of selecting out symmetry group singlets. However, a correction term is sometimes required to generate Hilbert series that correctly capture a full finite and small set of  lowest dimension  operators ($\Delta \le d$, in $d$ spacetime dimensions) of a particular form. This happens if there are operators encoded in $Z(\Phi, q, x, y)$ that are co-closed but not co-exact forms $\cO^{[\mu_1 \cdots \mu_k]}$:
\begin{equation}
\partial_{\mu_1} \cO^{[\mu_1 \cdots \mu_k]} = 0
\qquad\text{while}\qquad
\cO^{[\mu_1 \cdots \mu_k]} \ne \partial_\nu \cO^{[\nu \mu_1 \cdots \mu_k]} \,.
\label{eqn:cohomology}
\end{equation}
The issue is that when such ``cohomology'' operators exist in $Z(\Phi, q, x, y)$, dividing by the momentum generating function $P(q,x)$ will not correctly mod out the integration by parts redundancy. A correction piece $\Delta\Hcal$ will be needed and the full Hilbert series will be given by
\begin{equation}
\Hcal (\Phi, q) = \Hcal_0 (\Phi, q) + \Delta \Hcal (\Phi, q) \,.
\label{eqn:HFull}
\end{equation}
To address this issue, we provide in \cref{tbl:CohomologyForms} our manual enumeration of a list of co-closed but not co-exact form operators $\cO^{[\mu_1 \cdots \mu_k]}$ that could exist in $Z(\Phi, q, x, y)$ for a generic class of fields $\Phi$ including scalars, fermions, field strengths, and linearized Goldstone fields (see \cref{subsec:pion} for definition). The corresponding contributions to $\Delta\Hcal$ are also given in case of considering no internal symmetries. When we further impose internal symmetries, the contribution to $\Delta\Hcal$ from each co-closed but not co-exact $k$-form $\cO^{[\mu_1 \cdots \mu_k]}$ should be computed as
\begin{equation}
\Delta \Hcal (\Phi, q) \supset (-1)^{k+1} q^{\Delta_\cO+k} \int\text{d}\mu_\text{Internal}^{}(y)\, \chi_\cO(\Phi,y) \,,
\label{eqn:DeltaH}
\end{equation}
where $\Delta_\cO$ is the mass dimension of the $k$-form $\cO^{[\mu_1 \cdots \mu_k]}$ and $\chi_\cO(\Phi,y)$ is its field-graded internal symmetry character. For a derivation of \cref{eqn:DeltaH}, we refer the reader to Section 7.2 in \cite{Henning:2017fpj}. Let us mention that if all the fields in an EFT form unitary representations of the conformal group (such as in SMEFT), then the correction piece $\Delta \Hcal$ can be systematically computed as explained in Section 4.1 (specifically, Eq. (4.16b)) in \cite{Henning:2017fpj}. On the other hand, if there are non-unitary conformal representations involved in the EFT, such as the linearized Goldstone field in HEFT, then one needs to use the manual enumeration of the ``cohomology'' forms $\cO^{[\mu_1 \cdots \mu_k]}$ in \cref{tbl:CohomologyForms} (lower section) together with the formula in \cref{eqn:DeltaH}.

In summary, the full Hilbert series can be computed as in \cref{eqn:HFull}, concretely through \cref{eqn:H0,eqn:DeltaH}. The key tasks are to work out the correct SPMs $R_{\Phi_i}$ (and their characters $\chi_{\Phi_i}^{}$) for computing the $H_0$ piece, and to find out all the co-closed but not co-exact forms $\cO^{[\mu_1 \cdots \mu_k]}$ (contained in $Z(\Phi, q, x, y)$) for computing the $\Delta \Hcal$ piece.

\section{Massive Single Particle Modules and Higgs Mechanism}
\label{sec:SPMs}

In this section, we discuss the single particle modules (SPMs) and their characters. These characters are single particle partition functions graded by mass dimension and spin, and are the key ingredients for working out the quantity $Z(\Phi, q, x, y)$ in \cref{eqn:H0}, as explained around \cref{eqn:Zdecompose}. We begin with a basic review in \cref{subsec:review} and move on to discuss SPMs for Goldstones in \cref{subsec:pion} and field strengths in \cref{subsec:massless}, which are relevant for non-linearly realized gauge theories, such as HEFT. In \cref{subsec:massive}, we construct the SPM for a massive vector boson and show that it is a direct sum of the field strength SPM and the Goldstone SPM---the Higgs mechanism. In Appendix~\ref{appsec:spm} we give a general treatment of massive particles in arbitrary spacetime dimension \(d\), detailing their mode decomposition and computing their associated characters.

\subsection{Notation and Review of SPMs}
\label{subsec:review}

Throughout this section we adopt the language and notation of Ref.~\cite{Henning:2017fpj}, to which we refer the reader for further background and elaboration.\footnote{In particular, see section 2 of~\cite{Henning:2017fpj} for the main idea of SPMs, and section 3 for character theory.} We work in \(d=4\) with Euclidean Lorentz group \(SO(4) \simeq SU(2)\times SU(2)\). Finite dimensional, irreducible representations (irreps) of \(SO(4)\) are labeled by partitions \(l = (l_1,l_2)\) with \(l_{1,2}\) half integers and \(l_1 \ge \abs{l_2}\).\footnote{We could also label these representations according to \(SU(2)\times SU(2)\) using the vector \((j_1,j_2)\), where \(j_1 = (l_1 + l_2)/2\) and \(j_2 = (l_1-l_2)/2\).} For example, the vector (defining) representation is \((l_1,l_2) = (1,0)\), the adjoint representation is the direct sum \((1,1)\oplus(1,-1)\), the left-handed Weyl fermion representation is \((\frac{1}{2},\frac{1}{2})\), \textit{etc}. We will denote an \(SO(4)\) representation space as \(V_l \equiv V_{(l_1,l_2)}\), or simply by \(l \equiv (l_1,l_2)\) when there is no chance for confusion.\footnote{On occasion, and especially in \cref{appsec:spm}, we also use Young diagrams to denote irreps, \eg\ \(\ytableausetup{boxsize=.5em,centertableaux} V_{(n,0)} \equiv (n,0) \equiv \ydiagram{2}\cdots \ydiagram{1}\).}

The main ingredient in constructing an operator basis is the single particle module (SPM) \cite{Henning:2017fpj} \(R_{\Phi}\) for each particle/field \(\Phi\) to be included in the EFT. The SPM contains the fundamental building blocks used to construct composite, local operators. In essence, the SPM consists of the modes of a field in a Taylor expansion\footnote{If the particle is a conformal representation, the modes in the Taylor expansion are in one-to-one correspondence with states via the operator-state correspondence. This correspondence morally holds in a free massive theory as well. The existence of a well-defined expansion can be made rigorous in a CFT, where the expansion takes place on conformally compactified Minkowski space. Specifically, the conformal representation is realized as an analytic function---guaranteeing the convergence of the Taylor expansion---living on complexified Minkowski space, specifically the tube domain \(z^{\m} = x^{\m} + i y^{\m}\), with \(y\) time-like and forward in time, \(y^2>0\) and \(y^0>0\). The Minkowski representation is obtained in the limit \(y \to 0\) (analogous to an \(i\epsilon\) prescription), where the representation gets realized as a distribution.}
\begin{equation}
\Ph(x) = \blue{\Ph(0)} + \blue{\pd_{\m}\Ph(0)}\, x^{\m} + \frac{1}{2!} \blue{\pd_{\m}\pd_{\n}\Ph(0)}\, x^{\m}x^\n + \cdots \,,
\label{eqn:Taylor_Phi}
\end{equation}
where we have highlighted in blue the modes which form the building blocks of the SPM. The field \(\Ph\) obeys equations of motion which ensure the correct transformation properties under Poincar\'e symmetry, e.g.~\cite{Weinberg:1995mt} (see also~\cite{Bekaert:2006py} for a treatment in general spacetime dimension). The equations of motion are differential equations, which we can collectively denote by \(\{L\cdot \Ph = 0\}\) where the \(L\) are linear differential operators. Crucially, any terms proportional to the EOM are \textit{absent} in the expansion \cref{eqn:Taylor_Phi}.

Let's consider the simplest example: a massless scalar field \(\ph(x)\). The equation of motion is \(\pd^2 \ph = 0\), giving the expansion
\begin{equation}
\ph(x) = \blue{\ph(0)} + \blue{\pd_{\m}\ph(0)}\, x^{\m} + \frac{1}{2!} \blue{\pd_{\{\m}\pd_{\n\}}\ph(0)}\, x^{\m}x^\n + \cdots \,,
\end{equation}
where the curly brackets denote taking the derivatives in a traceless combination. Each mode \(\pd_{\{\m_1}\cdots \pd_{\m_n\}}\ph\) transforms in the \(SO(4)\) irrep \(V_{(n,0)}\), so that the SPM \(R_\ph\) is given by
\begin{equation}
R_{\ph} = \begin{pmatrix} \ph \\ \pd_{\m}\ph \\ \pd_{\{\m_1}\pd_{\m_2\}}\ph \\ \vdots \end{pmatrix} ~~\Rightarrow~~ R_{\ph} = \bigoplus_{n=0}^{\infty}V_{(n,0)} \,.
\label{eqn:scalar_SPM}
\end{equation}

As reviewed in \cref{sec:Recap}, when computing the Hilbert series a major set of input ingredients are the characters \(\ch_{\Ph_i}^{}\) for each SPM \(R_{\Ph_i}\). These characters are sums of the \(SO(4)\) characters \(\ch^{}_{(l_1,l_2)}(x)\) for each mode in the SPM, weighted by the scaling dimension of the mode. Here \(x = (x_1,x_2)\) are variables parameterizing the torus \(U(1)^2 \subset SO(4)\). Using the variable \(q\) as the weight for scaling dimension, and recalling \([\ph] = 1\) in \(d=4\), the character for the scalar field \(R_\ph\) in \cref{eqn:scalar_SPM} is given by
\begin{equation}
\ch_{\ph}(q,x) = \sum_{n=0}^{\infty}q^{n+1}\ch_{(n,0)}(x) \,.
\end{equation}
This can be summed directly, see \eg\ \cite{Henning:2017fpj}, to give
\begin{equation}
\ch_{\ph}(q,x) = q\,(1-q^2)\, P(q,x) \,,
\label{eqn:scalar_char}
\end{equation}
where the ``momentum generating function'' \(P(q,x)\) is given by
\begin{equation}
P(q,x) \equiv \frac{1}{\text{det}_{(1,0)}(1 - q g)} = \frac{1}{(1-qx_1)(1-q/x_1)(1-qx_2)(1-q/x_2)} \,.
\label{eqn:Pqx}
\end{equation}
We call this a momentum generating function because it is the character function for an infinite tower of derivatives
\begin{equation}
\begin{pmatrix}1 \\  \pd_\m \\ \pd_{\m_1}\pd_{\m_2} \\ \vdots \end{pmatrix} = \bigoplus_{n=0}\text{sym}^n(V_{(1,0)}) ~~\Rightarrow~~\sum_{n=0}q^n\text{sym}^n[\ch_{(1,0)}(x)] = P(q,x) \,.
\label{eqn:deriv_tower}
\end{equation}
This provides a useful way of understanding \cref{eqn:scalar_char}. In \(R_\ph\), \cref{eqn:scalar_SPM}, the derivatives are \textit{traceless} because of the EOM \(\pd^2\ph = 0\). One can then think of \(R_\ph\) schematically as
\begin{equation}
R_{\ph} = \begin{pmatrix} \ph \\ \pd_{\m}\ph \\ \pd_{\{\m_1}\pd_{\m_2\}}\ph \\ \vdots \end{pmatrix} \sim \begin{pmatrix} 1 \\ \pd_{\m} \\ \pd_{\m_1}\pd_{\m_2} \\ \vdots \end{pmatrix} (1- \pd^2)\, \ph \,.
\label{eqn:scalar_SPM_schematic}
\end{equation}
Combined with \cref{eqn:deriv_tower}, this makes the character \(\ch_\ph\) in \cref{eqn:scalar_char} completely manifest. This heuristic understanding of the character will be useful for quickly deriving the character of a massive vector.

\subsection{Pion SPM in Nonlinear Realizations}
\label{subsec:pion}

In nonlinear realizations, the pion fields in the Goldstone matrix
\begin{equation}
\xi = e^{i\pi^a t^a/v} \,,
\end{equation}
transform nonlinearly under the symmetry group. We follow the CCWZ prescription \cite{Coleman:1969sm,Callan:1969sn} to use the Cartan-Maurer linearization variable\footnote{For a coset space \(G/H\), the Cartan-Maurer form \(\xi^{-1}d\xi\) is valued in the Lie algebra \(\mathfrak{g}\simeq \mathfrak{g}/\mathfrak{h} \oplus \mathfrak{h}\) of \(G\). The notation \(\xi^{-1}d\xi|_{\text{coset}}^{}\) means to take the piece valued in \(\mathfrak{g}/\mathfrak{h}\), i.e. the piece proportional to the broken generators.} 
\begin{equation}
u_\mu = u_\mu^a t^a \equiv \Big[ \xi^{-1} \left( \partial_\mu \xi \right) \Big] \Big|_\text{coset} \,,
\end{equation}
as the building block of the EFT Lagrangian, which transforms linearly under the unbroken group.

To work out the SPM $R_u$, let us first consider the simplest case that the pion coset space is $U(1)$. In this case, we simply have
\begin{equation}
\xi = e^{i\pi/v} \,,\qquad
u_\mu = \frac{i}{v}\, \partial_\mu \pi \,,
\end{equation}
and the SPM is given by
\begin{equation}
R_u = R_{\pd\pi} = \frac{i}{v} \begin{pmatrix} \pd_{\m}\pi \\ \pd_{\{\m_1}\pd_{\m_2\}}\pi \\ \vdots \end{pmatrix} ~~\Rightarrow~~ R_{\pd\pi} = \bigoplus_{n=0}^{\infty}V_{(n+1,0)} \,.
\label{eqn:pion_SPM}
\end{equation}
Comparing with \(R_\ph\) in \cref{eqn:scalar_SPM}, we see that the only difference is the absence of the scalar mode \(\pi\) (with no derivatives). This can be understood intuitively by the softness requirement---pion amplitudes vanish when the momentum vanishes---which leads to a derivative expansion in the effective Lagrangian. Taking the $u_\mu$ field to be mass dimension one, \([u_\mu] = [\frac{i}{v}\, \partial_\mu\pi] = 1\), the corresponding character is given by
\begin{equation}
\ch_u (q,x) = \ch_{\pd \pi}^{} (q,x) = \sum_{n=0}^{\infty} q^{n+1}\ch_{(n+1,0)}(x) = (1-q^2)P(q,x) - 1 = \frac{1}{q}\ch_{\ph}(q,x) - 1 \,.
\label{eqn:pion_char}
\end{equation}

In the above, we have derived the SPM $R_u$ and its character $\chi_u$ by considering the simplest coset space $U(1)$. But the results in \cref{eqn:pion_SPM,eqn:pion_char} actually hold for general coset spaces~\cite{Henning:2017fpj}. Of course, in general cases one should make the replacement
\begin{equation}
\frac{i}{v}\, \partial_\mu \pi \quad\longrightarrow\quad
u_\mu = \frac{i}{v}\, \pd_{\m}\pi + O\left(\pi^2\right) \,,
\end{equation}
in \cref{eqn:pion_SPM}, and the SPM reads
\begin{equation}
R_u = \begin{pmatrix} u_\mu \\ \pd_{\{\m_1} u_{\mu_2\}} \\ \vdots \end{pmatrix} \,.
\end{equation}
Nevertheless, the right panel of \cref{eqn:pion_SPM} still holds, thanks to the following properties of the field $u_\mu$ (see CCWZ \cite{Coleman:1969sm,Callan:1969sn}, as well as Section 7 of \cite{Henning:2017fpj} for details):
\begin{subequations}
\begin{alignat}{2}
\text{EOM}:\qquad && \partial^\mu u_\mu &= 0 \,, \\[5pt]
\text{Vanishing Curl}:\qquad && \partial_{[\mu} u_{\nu]} &= 0 \,.
\end{alignat}
\end{subequations}
Here the square brackets refer to the anti-symmetric combination of indices. Therefore, the character result in \cref{eqn:pion_char} holds for general coset spaces \(G/H\) associated with the breaking of internal symmetries, i.e. \(G\) and \(H\) are compact, semi-simple groups.

\subsection{Massless Vector SPM}
\label{subsec:massless}

The interpolating field for a massless vector boson is the field strength \(F_{\m\n}(x)\). In \(d=4\), it can be split into its chiral components \(F_{L/R}\). The corresponding SPMs are
\begin{subequations}\label{eqn:gauge_SPM}
\begin{align}
R_{F_L} &= \bigoplus_{n=0}^{\infty}V_{(n+1,1)} \,, \\[8pt]
R_{F_R} &= \bigoplus_{n=0}^{\infty}V_{(n+1,-1)} \,, \\[8pt]
R_{F} &= R_{F_L}\oplus R_{F_R} \,,
\end{align}
\end{subequations}
with corresponding characters
\begin{subequations}\label{eqn:gauge_char}
\begin{align}
\ch^{}_{F_L}(q,x_1,x_2) &= \sum_{n=0}^{\infty}q^{n+2}\ch_{(n+1,1)}(x_1,x_2) \nonumber \\[5pt]
&= q^2\Big(\ch_{(1,1)}(x_1,x_2) - q \ch_{(1,0)}(x_1,x_2) + q^2\Big)P(q,x_1,x_2) \,, \\[8pt]
\ch^{}_{F_R}(q,x_1,x_2) &= \sum_{n=0}^{\infty}q^{n+2}\ch_{(n+1,-1)}(x_1,x_2) \nonumber \\[5pt]
&= q^2\Big(\ch_{(1,-1)}(x_1,x_2) - q \ch_{(1,0)}(x_1,x_2) + q^2\Big)P(q,x_1,x_2) \,.
\end{align}
\end{subequations}

\subsection{Massive Vector SPM}
\label{subsec:massive}

A massive vector \(A_{\mu}\) is described by the Proca equations
\begin{subequations}\label{eqn:proca}
\begin{align}
(\pd^2 + m^2)A_{\m} &=0 \,, \label{eqn:proca_KG} \\[5pt]
\pd^\mu A_\mu &= 0 \,. \label{eqn:proca_transverse}
\end{align}
\end{subequations}
The first of these is the Klein-Gordon equation, dictating that the particle has mass \(m\). The second of these is a polarization condition, which in momentum space translates to \(p^{\m} \e^{\s}_\m(p) = 0\) where \(\e^\s_\m(p)\) is the polarization tensor. In analogy with the heuristic scalar SPM in \cref{eqn:scalar_SPM_schematic}, we see that the SPM for \(A_\m\)  has the  following heuristic form 
\begin{equation}
R_{A} \sim \begin{pmatrix} 1 \\ \pd_{\m} \\ \pd_{\m_1}\pd_{\m_2} \\ \vdots \end{pmatrix} (1- \pd^2)(1-\pd^{\m})A_{\m} \,,
\label{eqn:vector_SPM_schematic}
\end{equation}
which leads us to the character
\begin{equation}
\ch_{A}^{} (q,x) = q\,(1-q^2)\, \Big(\ch^{}_{(1,0)}(x) - q\Big)P(q,x) \,.
\label{eqn:A_char}
\end{equation}
The \((1-q^2)\) factor is associated to the Klein-Gordon equation~\eqref{eqn:proca_KG}, analogous to the scalar case in Eq.~\eqref{eqn:scalar_SPM_schematic}, while the \((\ch^{}_{(1,0)} - q)\) reflects Eq.~\eqref{eqn:proca_transverse}: \(A_\m\) is a vector (hence \(\ch_{(1,0)}\)) while \(\pd^\m A_\m\) is a scalar and involves a derivative (hence \(q\, \ch^{}_{(0,0)} = q\)). Some insight into this character---and strong evidence that we have indeed constructed the right quantity---is obtained by recognizing that it is the sum of the characters for a Goldstone, $\chi_u$ in \cref{eqn:pion_char} and a massless gauge field, \(\ch_F^{} \equiv \ch_{F_L}^{} + \ch_{F_R}^{}\) in \cref{eqn:gauge_char}:
\begin{equation}
\ch_A^{} (q,x) = \ch_F^{} (q,x) + \ch_u (q,x) \,.
\label{eqn:HiggsMechanism}
\end{equation}
This decomposition reflects the Higgs mechanism: a massive vector can be thought of as the combination of transverse modes \(\vec{p}\cdot \vec{\e}^{\, \s} = 0\)---interpolated by a field strength \(F_{\m\n}\)---together with the longitudinal mode \(\vec{p}\cdot \vec{\epsilon}^{\, \s} \ne 0\)---interpolated by a Goldstone field \(u_\mu\).

\begin{figure}[t]
\begin{center}
  \begin{tikzpicture}[>=latex]
    \node (A) at (0,0) {\(\def\arraystretch{1.8} \ytableausetup{boxsize=.6em,centertableaux}
    \begin{array}{rccccccc}
      \text{massless spin-}0: & ~~1~~ & ~~\ydiagram{1}~~ & ~~\ydiagram{2}~~ & ~~\ydiagram{3}~~ & ~~\ydiagram{4}~~ & ~~\ydiagram{5}~~ & \cdots \\
      \text{massless spin-}1: & & \ydiagram{1,1} & \ydiagram{2,1} & \ydiagram{3,1} & \ydiagram{4,1} & \ydiagram{5,1} & \cdots \\
      \text{massless spin-}2: & & & \ydiagram{2,2} & \ydiagram{3,2} & \ydiagram{4,2} & \ydiagram{5,2} & \cdots \\
      \text{massless spin-}3: & & & & \ydiagram{3,3} & \ydiagram{4,3} & \ydiagram{5,3} & \cdots \\
      \text{massless spin-}4: & & & & & \ydiagram{4,4} & \ydiagram{5,4} & \cdots 
    \end{array}
    \)};
    
    \draw[-,blue,thick] (-2.7,2.1) to (6.5,2.1);
    \draw[-,blue,thick] (-2.7,2.1) to (-2.7,1.45);
    \draw[-,blue,thick] (-2.7,1.45) to (6.5,1.45);
    
    \draw[-,magenta,thick] (-2,.4) to (6.5,.4);
    \draw[-,magenta,thick] (-2,.4) to (-2,2.15);
    \draw[-,magenta,thick] (-2,2.15) to (6.5,2.15);
    
    \draw[-,cyan,thick] (-1.1,-.6) to (6.5,-.6);
    \draw[-,cyan,thick] (-1.1,-.6) to (-1.1,2.2);
    \draw[-,cyan,thick] (-1.1,2.2) to (6.5,2.2);

    \draw[<-,blue,thick] (-2.5,2.2) to [out=60,in=180] (-1.5,2.6);
    \node at (-.2,2.6) {\textcolor{blue}{massive spin-0}};

    \draw[<-,magenta,thick] (-1.7,.3) to [out= 210,in=90] (-2.4,-.9);
    \node at (-1.4,-1.1) {\textcolor{magenta}{massive spin-1}};

    \draw[<-,cyan,thick] (.1,-.7) to (.1,-2.0);
    \node at (-.8,-2.2) {\textcolor{cyan}{massive spin-2}};    

  \end{tikzpicture}
\end{center}
\caption{The mode decomposition of massive single particle modules reflects the Higgs mechanism: a massive particle can be thought of as the combination of transverse polarizations---interpolated by a massless particle of the same spin---together with longitudinal polarizations that are interpolated by a series of massless particles of lower spin, e.g. Eqs.~\eqref{eqn:RADecomposition} and~\eqref{eqn:HiggsMechanism} for a massive vector, and Eq.~\eqref{eq:MassiveSpinkSPMdecomp} for general massive spin-\(k\) particles. For the treatment of massive particles of general spin see App.~\ref{appsec:spm}.}
\label{fig:Higgs_Mechanism}
\end{figure}
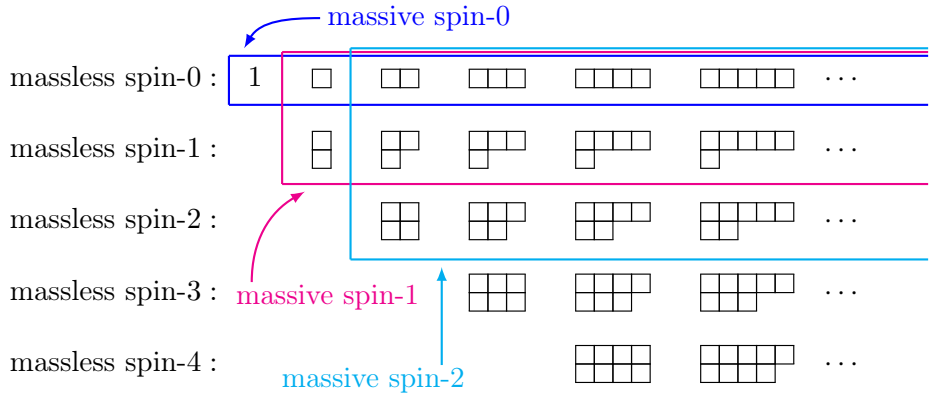

One could also carefully derive \cref{eqn:HiggsMechanism} (and then \cref{eqn:A_char} consequently) by considering the representation decomposition for each block (row) in $R_A$, the same procedure shown in \cref{eqn:scalar_SPM,eqn:pion_SPM,eqn:gauge_SPM}. To this end, we first note the following symmetric + anti-symmetric tensor decomposition
\begin{equation}
\pd_{\mu_1} \cdots \pd_{\mu_n} A_\nu = \pd_{\mu_1} \cdots \pd_{\mu_{n-1}} \pd_{\{\mu_n} A_{\nu\}} \;\oplus\; \pd_{\mu_1} \cdots \pd_{\mu_{n-1}} \pd_{[\mu_n} A_{\nu]} \,,
\end{equation}
which is relevant for the $(n+1)$-th block (schematically $A\,\pd^n$) in $R_A$. Now further taking into account of the conditions in \cref{eqn:proca} for $R_A$, it is clear that the symmetric components in $R_A$ will form $R_u$ and the anti-symmetric ones will form $R_F$:
\begin{equation}
R_A = R_u \oplus R_F \,,
\label{eqn:RADecomposition}
\end{equation}
and therefore \cref{eqn:HiggsMechanism} follows.

In fact, the ``Higgs mechanism'' reflected in \cref{eqn:RADecomposition} generalizes to massive particles of higher spins. Decomposing the polarizations of a massive particle into transverse and longitudinal components, the various longitudinal polarizations can be thought of originating from massless particles of lower spin that the massive particle has ``eaten''. We elaborate on this in \cref{appsec:spm}, where we extend our treatment to massive particles of general spin. The basic result, however, is simple to understand and can be gleaned from \cref{fig:Higgs_Mechanism}, which shows how a massive spin-\(k\) SPM decomposes into massless particles of spin \(l = k, k-1,\dots,0\).

Finally we comment on various grading options available when using these massive characters inside the Hilbert series calculation. As reviewed in~\cref{eqn:Zdecompose}, the character \(\ch_{\Ph}^{}\) for the SPM \(R_{\Ph}\) is typically multiplied by a grading variable \(\Ph\) when inserted into the plethystic exponential, allowing us to keep track of where the field \(\Ph\) shows up. So, for example, in including a massive vector \(A_{\m}\) we could take the massive vector character~\cref{eqn:A_char} and multiply it by a grading variable \(A\): \(\text{PE}[A\,\ch_A^{}(q,x_i)]\). However, such a choice is not mandatory, and in the case that the character entering the SPM has some nice decomposition---\textit{e.g.} like the Higgs mechanism reflected in~\cref{eqn:RADecomposition} and~\cref{eqn:HiggsMechanism}---it can be very convenient to grade the pieces separately. For the case of~\cref{eqn:HiggsMechanism}, for example, we could include separate grading variables for the transverse and longitudinal pieces, \(\text{PE}[A\, \ch_{A}^{}(q,x_i)] \to \text{PE}[F\, \ch_F^{}(q,x_i)+u\, \ch_u(q,x_i)]=\text{PE}[F\,\ch_F^{}(q,x_i)]\text{PE}[u\,\ch_u(q,x_i)]\). In such a case, terms in the Hilbert series containing the grading variable \(F\) are built out of the field strength \(F_{\m\n}\) while those containing \(u\) are built out of the Cartan-Maurer linearization variable (or, in unitary gauge, built with \(A_{\m}\) directly). We find such a splitting useful in our analysis of HEFT in~\cref{sec:hefths}.

\section{Form Factor Redundancies and Spurion Fields}
\label{sec:Spurion}

In this section, we discuss how to mod out a new type of redundancy relation in the Hilbert series---singlet polynomial factors of a certain field $T$:
\begin{equation}
\cO_1 \sim \cO_2 \,,\quad\text{if}\quad
\cO_1 = f(T)\, \cO_2
\quad\text{or}\quad
\cO_2 = f(T)\, \cO_1 \,,
\label{eqn:FFredundancy}
\end{equation}
where $f(T)$ is a polynomial of $T$ that is a singlet under the EFT symmetries. This redundancy relation is ``new'' in the sense that it is not in the set of redundancies that one usually accounts for in constructing EFT operator bases---equations of motion, integration by parts, and group identities. It is motivated by the observation that in some EFTs, singlet polynomials of a certain field $T$ are wrapped into form factors in the definition of the operator basis, which amounts to imposing the above redundancy relation. For example, this is how the physical Higgs field $h$ is treated in HEFT. Another motivation for considering the redundancy in \cref{eqn:FFredundancy} is that it is also encountered when we introduce spurion fields to classify the operators into preserving/breaking sets of a bigger symmetry group, where singlet polynomials of the spurion fields are trivial factors yielding a redundancy relation of the above kind in the full Hilbert series. This also finds applications in the case of HEFT.

To handle the redundancy relation in \cref{eqn:FFredundancy}, the key fact we make use of is a factorization property of Hilbert series. Consider the polynomial space of two independent set of variables $T=\{T_1, T_2, \cdots \}$ and $\phi=\{\phi_1, \phi_2, \cdots\}$. These variables form representations of certain symmetry groups, and we are selecting out a subspace of their polynomials that preserve the symmetries. Let us use the Hilbert series $\Hcal_\text{All} (T,\phi)$ to encode this subspace, then it has the factorization property
\begin{equation}
\Hcal_\text{All} (T,\phi) = \Hcal_\text{All} (T,\phi=0)\; \Hcal_{T\text{-Quotient}} (T, \phi) \,.
\label{eqn:Factorization}
\end{equation}
Here $\Hcal_\text{All} (T,\phi)$ encodes all the singlet polynomials, so by definition $\Hcal_\text{All} (T,\phi=0)$ will encode all the singlet polynomials made of $T$ alone. The point is that the remaining factor $\Hcal_{T\text{-Quotient}} (T, \phi)$ corresponds to a well-defined polynomial space $\{\cO_a(T,\phi)\}$, where elements are not related by a (singlet) polynomial factor of $T$:
\begin{equation}
\cO_a (T,\phi) \ne f(T)\, \cO_b (T,\phi) \,.
\end{equation}
Therefore, $\Hcal_{T\text{-Quotient}} (T, \phi)$ encodes the space of operators with singlet polynomial factors of $T$ modded out. This is precisely what we are looking for. From \cref{eqn:Factorization}, we see that the Hilbert series for the quotient space can be computed as
\begin{equation}
\Hcal_{T\text{-Quotient}} (T, \phi) = \frac{\Hcal_\text{All} (T,\phi)}{\Hcal_\text{All} (T,\phi=0)} \,.
\label{eqn:Quotient}
\end{equation}

It is not hard for one to convince themselves that \cref{eqn:Factorization} is true. In terms of the operator space, it is simply reflecting the following relation
\begin{align}
\Big\{ \text{All singlet operators} \Big\} &= \Big\{\text{singlets of } T \text{ alone: } f(T) \Big\} \notag\\[5pt]
&\hspace{20pt}
\otimes \Big\{ \cO_a(T,\phi), \text{with } \cO_a (T,\phi) \ne f(T)\, \cO_b (T,\phi)\Big\} \,.
\label{eqn:FactorizationOps}
\end{align}
The only nontrivial check of this relation is perhaps that there is no double counting on the right-hand side, which is true because of the fundamental theorem of arithmetic.

\subsection{Modding out Form Factor Redundancies}
\label{sec:Spuriona}

In HEFT, it is customary to group powers of the physics Higgs field $h$ into form factors and write each independent operator as  $f(h)\, \cO$, with $f(h)$ an arbitrary polynomial of $h$. This amounts to imposing the redundancy relation in \cref{eqn:FFredundancy} with $T=h$, so we can make use of \cref{eqn:Quotient} to mod it out and obtain the Hilbert series for the quotient space. As $T=h$ is a singlet under the HEFT symmetries, we have
\begin{equation}
\Hcal_\text{All} (h,\phi=0) = \frac{1}{1-h} \,.
\end{equation}
and therefore
\begin{equation}
\Hcal_{h\text{-Quotient}} (h, \phi) = (1-h)\, \Hcal_\text{All} (h,\phi) \,.
\label{eqn:hFF}
\end{equation}
Note that $\phi$ here collectively denotes all the other field components in HEFT that we are not wrapping into form factors; in particular, it contains the derivatives of $h$.

One could also imagine defining a new type of operator basis for SMEFT where singlet polynomials of the Higgs doublet $H$ is wrapped into form factors (as what is hinted at in \eg\ \cite{Helset:2020yio}). In such a case, the field $H$ is not a singlet itself, so we have a relatively nontrivial factor
\begin{equation}
\Hcal_\text{All} (H,\phi=0) = \frac{1}{1-H^\dagger H} 
\quad\longrightarrow\quad
\Hcal_{H\text{-Quotient}} (H, \phi) = (1-H^\dagger H)\, \Hcal_\text{All} (H,\phi) \,.
\end{equation}

\subsection{Hilbert Series with Spurion Fields}
\label{sec:Spurionb}

Consider the situation where the field $\phi$ forms a representation of a group $G$, but we only require the operators to preserve a subgroup of it $H\subset G$. In this case, we may take two approaches to obtain the Hilbert series, which will of course give us the same result:
\begin{enumerate}
\item \textbf{Direct Approach:} We write out the (reducible) representation of $\phi$ under the subgroup $H$ and then compute the Hilbert series by directly requiring $H$ invariance. This is arguably the most direct way of attacking the problem, and let us denote the resulting Hilbert series as $\mathcal{H}_H(\phi)$. 
\item \textbf{Spurion Approach:} We introduce a spurion field $T$ for assistance. Its representation under the larger group $G$ is chosen such that upon taking a \emph{generic} vev, it breaks $G$ down to $H$. We can then compute the Hilbert series with $\phi$ and $T$ together, requiring $G$ invariance. We denote the Hilbert series obtained this way as $\mathcal{H}_G(T, \phi)$. The spirit of this approach is that although we seem to be preserving a larger symmetry group $G$, eventually we will be just enumerating $H$-preserving operators, because we will send the spurion to its vev value. This approach might sound like a detour, but it is useful because it allows us to classify the $H$-preserving operators into $G$-preserving/breaking sectors, \ie\ terms in $\mathcal{H}_G(T, \phi)$ without/with $T$. Indeed in HEFT a spurion \textbf{T} is often introduced that breaks the custodial symmetry $G=SU(2)_V$ down to $H=U(1)_\text{EM}$ (for the Higgs sector). However, $\mathcal{H}_G(T, \phi)$ as described above actually encodes a lot more operators out of our interests, because there are singlet factors made of the spurion field $T$ alone, each of which could be multiplying the informative ones that involve $\phi$. Clearly, this is a redundancy of the type in \cref{eqn:FFredundancy} and we can make use of \cref{eqn:Quotient} to mod it out. So what we actually need is
\begin{equation}
\Hcal_{G\text{-Classified}} (T, \phi) = \frac{\mathcal{H}_G (T, \phi)}{\mathcal{H}_G (T, \phi=0)} \,.
\label{eqn:HClassified}
\end{equation}
Then the $\mathcal{H}_H (\phi)$ in the direct approach can be reproduced by dropping the $G$-preserving/breaking classification information, \ie\ by taking $T \to 1$:
\begin{equation}
\mathcal{H}_H (\phi) = \lim_{T\to 1} \Hcal_{G\text{-Classified}} (T, \phi) = \lim_{T\to 1} \frac{\mathcal{H}_G (T, \phi)}{\mathcal{H}_G (T, \phi=0)} \,.
\label{eqn:Equivalence}
\end{equation}
\end{enumerate}

In the rest of this subsection, we show a few simple examples to demonstrate the equivalence between the above two approaches, in particular, the relation in \cref{eqn:Equivalence}. For simplicity, we only consider operators with no derivatives in these examples.

\subsubsection{Example: A $U(1)$ Spurion}
\label{subsubsec:U1spurion}

Let us first consider an example with $G=U(1)=\{e^{i\theta}, \theta\in [0,2\pi]\}$ broken down to the cyclic subgroup $H=\mathbb{Z}_n$ with $\{\theta=2\pi\frac{k}{n}, k=0,1,\cdots,n-1\}$. Our field is a complex scalar $\phi$ with charge $+1$ under $G=U(1)$; its complex conjugate $\phi^*$ of course has charge $-1$:
\begin{equation}
\phi \to \tilde\phi = e^{i\theta} \phi
\qquad,\qquad
\phi^* \to \tilde\phi^* = e^{-i\theta} \phi^* \,.
\end{equation}

We use the direct approach first. The invariants of the unbroken group $H=\mathbb{Z}_n$ are generated by
\begin{equation}
I_1=\phi^*\phi \,,\quad 
I_2=\phi^n \,,\quad
I_3=\left(\phi^*\right)^n \,.
\end{equation}
But these are not free generators, as there is one redundancy relation among them
\begin{equation}
I_1^n = I_2 I_3 \,.
\end{equation}
Therefore, following the general understanding (\eg\ \cite{Jenkins:2009dy}), we obtain the Hilbert series
\begin{equation}
  \mathcal{H}_H(\phi,\phi^*) = \frac{1- \phi^n\left(\phi^*\right)^n}{ \left(1-\phi^*\phi\right) \left[ 1-\phi^n \right] \left[1-\left(\phi^*\right)^n \right]} \,.
\label{eqn:HSU1Direct}
\end{equation}

Now let us use the spurion approach and show that it yields the same result via \cref{eqn:Equivalence}. We introduce a spurion $T$ which is also a complex scalar but with a charge assignment $q_T=n$ under $G=U(1)$. So it transforms as
\begin{equation}
T \to \tilde{T} = e^{i n \theta} T \,.
\end{equation}
Note that $T$ is unchanged under the $\mathbb{Z}_n$ subgroup. Therefore, when it obtains a vev, the unbroken subgroup is $\mathbb{Z}_n$, as we need. With our spurion $T$, we get the Hilbert series
\begin{equation}
\mathcal{H}_G(T,T^*,\phi,\phi^*) = \frac{1- \left(\phi^*\phi\right)^n T^*T}{\left(1-\phi^*\phi\right) \left(1-T^*T\right) \left[ 1-\phi^n T^* \right] \left[ 1-\left(\phi^*\right)^n T \right]} \,.
\end{equation}
We see that making use of \cref{eqn:Equivalence} does reproduce \cref{eqn:HSU1Direct}.

\subsubsection{Example: An $SO(N)$ Vector Representation Spurion}
\label{subsubsec:SONspurion}

Next we consider a real scalar field $\phi$ with $N\ge3$ real components that forms a vector representation of the group $G=SO(N)$, but we only impose the invariance under the subgroup $H=SO(N-1)$. The case of $N=3$ corresponds precisely to the case of custodial symmetry breaking in HEFT.

Again, let us use the direct approach first. Under the unbroken group $H=SO(N-1)$, the $N$ components of field $\phi$ form a vector representation and a singlet representation. Clearly, the Hilbert series is
\begin{equation}
\mathcal{H}_H (\phi) = \frac{1}{(1-\phi^2)(1-\phi)} \,.
\label{eqn:HSSONDirect}
\end{equation}

Now let us use the spurion approach. We introduce a spurion $T$ that forms a vector representation of $G=SO(N)$, which will break $G$ down to $H=SO(N-1)$, as we need. With the spurion $T$, we will obtain the Hilbert series\footnote{For $N\ge3$, $(\vec\phi\cdot\vec\phi)$, $(\vec T \cdot \vec T)$, and $(\vec\phi \cdot \vec T)$ are free generators.}
\begin{equation}
\mathcal{H}_G (T, \phi) = \frac{1}{ \left(1-\phi^2\right) \left(1-T^2\right) \left(1-\phi T\right) } \,.
\end{equation}
Again, we see that making use of \cref{eqn:Equivalence} would then reproduce \cref{eqn:HSSONDirect}.

\section{Hilbert Series for ($\nu$)HEFT}
\label{sec:hefths}

A suitable and phenomenologically attractive effective field theory that can benefit from the techniques described and discussed in previous sections is the so called Higgs Effective Field Theory (HEFT). The motivation behind HEFT stems from the fact that a variety of BSM models with the Higgs not occupying an elementary exact $SU(2)_L$ doublet is still experimentally allowed; therefore, it is desirable to identify observables that would allow for distinguishing among the possible scenarios. To this end, a model independent approach represented by HEFT provides a very convenient framework. While in the SMEFT the physical Higgs and the three electroweak Goldstone bosons are incorporated in an electroweak doublet, in HEFT they are independent fields with the physical Higgs field $h$ being a SM gauge singlet. We also supplement the SM field content with a right-handed neutrino $\nu_R$, and call this EFT as ``$\nu$HEFT'' (in analogy to ``$\nu$SMEFT''~\cite{delAguila:2008ir,Bhattacharya:2015vja,Liao:2016qyd}).

We begin by summarizing the various technical aspects we have detailed above, in addition to those appearing in previous papers, that are relevant for applying Hilbert series techniques  to HEFT.

\begin{itemize}
\item \textbf{SPMs of the Goldstone bosons $R_u$:} In HEFT, the electroweak (EW) symmetry breaking is realized nonlinearly by the Goldstone matrix
\[
\xi = e^{i\pi^a t^a/v} \,.
\]
To apply the Hilbert series method, one needs to follow the CCWZ prescription \cite{Coleman:1969sm,Callan:1969sn} to use the linearized building blocks
\[
u_\mu = u_\mu^a t^a \equiv \Big[ \xi^{-1} \left( \partial_\mu \xi \right) \Big] \Big|_\text{coset} \,,
\]
which transform linearly under the unbroken group. We provided a brief review of the SPM for $u_\mu$ in \cref{subsec:pion}. Further details can be found in Section 7 of \cite{Henning:2017fpj}.
\item \textbf{SPMs of the gauge bosons $R_F$:} In HEFT, we are spontaneously breaking a symmetry group $G$ that is gauged. So in addition to the linearized Goldstone boson fields $u_\mu$, the field strengths $F_{\mu\nu}$ for the symmetry $G$ should also be included into the particle content. These field strengths $F_{\mu\nu}$ are dynamic fields of the EFT (as opposed to background fields), so they are also subject to the EOM redundancies (\ie\ the on-shell conditions). Note that this is different from the QCD Chiral Lagrangian case \cite{Graf:2020yxt} where the field strength are external fields that are not subject to the EOM redundancies. In \cref{subsec:massless}, we reviewed the SPM and character for dynamic field strengths.
\item \textbf{Gauge choices and the Higgs mechanism:} In HEFT, we are dealing with a gauged symmetry that is spontaneously broken. In this case, there are different gauge choices to describe the same physics. For example, one could choose the unitary gauge to eliminate the Goldstone bosons, and the dynamic fields would be the massive gauge bosons $A_\mu$, leading us to use the massive vector SPMs $R_A$ given in \cref{subsec:massive}. Equivalently one could use the modules $R_F$ and $R_u$, the equivalence being manifested by the Higgs mechanism relation:
\begin{equation}
R_A = R_F \oplus R_u \,,
\end{equation}
as explained in \cref{subsec:massive}.
\item \textbf{Non-unitary representations of the conformal group:} In HEFT, the linearized Goldstone building blocks $R_u$ do not form unitary representations of the conformal group. Because of this, the correction piece $\Delta \Hcal$ in \cref{eqn:HFull} cannot be computed using conformal representation theory (specifically, Eq. (4.16b) in \cite{Henning:2017fpj}). Instead, one needs to follow our manual enumeration of co-closed but not co-exact forms $\cO_{[\mu_1 \cdots \mu_k]}$ in \cref{tbl:CohomologyForms}, and then use \cref{eqn:DeltaH} to compute $\Delta \Hcal$. We will carry this out explicitly for HEFT in the below.
\item \textbf{Form factor equivalence relation:} --- When discussing HEFT operators it is customary to wrap powers of the singlet physical Higgs field $h$ into form factors and view
\[
f(h)\, \cO \,,\qquad\text{with $f(h)$ an arbitrary polynomial of $h$}\,,
\]
as a single operator. This amounts to introducing a new type of redundancy relation
\begin{equation}
\cO_1 \sim \cO_2 \,,
\quad\text{if}\quad
\cO_1 = f(h)\, \cO_2
\quad\text{or}\quad
\cO_2 = f(h)\, \cO_1 \,.
\end{equation}
Therefore, after obtaining the usual Hilbert series, we need to further mod out this additional redundancy. The technique to handle this was given in \cref{sec:Spuriona}.
\item \textbf{Custodial spurion:} Another customary practice in HEFT is to write the operators in the form of custodial $SU(2)_V$ invariants, despite the fact that the operators are only required to preserve $U(1)_\text{EM}$. Typically, a custodial symmetry breaking spurion field
\[
T \sim \sigma^3
\]
is introduced to restore the symmetry, \eg\ \cite{Brivio:2016fzo,Buchalla:2013rka,Krause:2016uhw,Sun:2022ssa} (see~\cref{eqn:Tdef} for the concrete expression). The resulting operators are in apparent custodial preserving combinations, but with various insertions of the spurion field $T$ to account for custodial breaking effects upon it taking a vev. Clearly, this is a less direct way of approaching the operator basis, but it should yield the same result for the $B-L$ preserving sectors. The additional information about the division into different powers of spurion $T$ might also be useful in some cases. Obtaining a Hilbert series that reflects this way of writing the operator basis is straightforward. One simply includes the spurion field $T$ into the particle content and imposes the larger symmetry $SU(2)_V$ instead of $U(1)_\text{EM}$. This technique was detailed in \cref{sec:Spurionb}.
\end{itemize}

\subsection{Nonlinear Realization of the Electroweak Symmetry Breaking}
\label{subsec:nonlinear}

($\nu$)HEFT is a nonlinear realization of the electroweak symmetry breaking
\begin{equation}
SU(2)_L \times U(1)_Y \quad\longrightarrow\quad
U(1)_\text{EM} \,.
\label{eqn:EWSB}
\end{equation}
It has an alternative name---the electroweak chiral Lagrangian (with a physical Higgs field $h$). Using $t^1, t^2, t^3$ and $Y$ to denote the $SU(2)_L \times U(1)_Y$ generators, we have
\begin{subequations}\label{eqn:Generators}
\begin{align}
\text{Broken Generators}\quad&
\left\{\begin{array}{l}
t^\pm = t^1 \pm i t^2 \\[3pt]
t^z = c_{\theta_W} \left( c_{\theta_W} t^3 \right) - s_{\theta_W} \left( s_{\theta_W} Y \right)
\end{array}
\right. \,, \\[5pt]
\text{Unbroken Generator}\quad&
c_{\theta_W} s_{\theta_W}\, Q = s_{\theta_W} \left( c_{\theta_W} t^3 \right) + c_{\theta_W} \left( s_{\theta_W} Y \right) \,,
\end{align}
\end{subequations}
where $\theta_W$ is the Weinberg angle. In nonlinear realization, each broken generator in \cref{eqn:Generators} is accompanied by a Goldstone field to form the Goldstone matrix
\begin{equation}
\xi = e^{i\left( \pi^+ t^+ + \pi^- t^- + \pi^z t^z \right)/v} \,,
\end{equation}
which transforms nonlinearly under the symmetry groups. One then follows the CCWZ prescription \cite{Coleman:1969sm,Callan:1969sn} to construct the Cartan-Maurer linearization variables
\begin{equation}
V_\mu = V_\mu^+\, t^+ + V_\mu^-\, t^- + V_\mu^z\, t^z \;\equiv\; \xi^{-1} \left(\partial_\mu \xi\right) \Big|_\text{coset} \,.
\label{eqn:Vmu}
\end{equation}
These linearized Goldstone fields are linear representations of the unbroken symmetry $U(1)_\text{EM}$. As suggested by the notation, $V_\mu^\pm, V_\mu^z$ correspond to the longitudinal modes of the gauge bosons $W^\pm_\mu, Z_\mu$ in light of the Higgs mechanism. In HEFT, other SM fields are also organized according to representations of the unbroken symmetry $U(1)_\text{EM}$. A summary of the field representations is given in \cref{tbl:FieldContent}. Note that the linearized Goldstone fields $V_\mu^\pm, V_\mu^z$ have canonical mass dimension one, because they contain one power of derivative; see \cref{eqn:Vmu}.

\begin{table}[t]
\renewcommand{\arraystretch}{1.3}
\setlength{\arrayrulewidth}{.2mm}
\setlength{\tabcolsep}{1em}
\centering
\begin{tabular}{cccccc}
\toprule
Field     & Lorentz Group                   & $SU(3)_C$                     & $U(1)_{\text{EM}}$
          & $\text{dim}$ \\
\midrule
$u_L \;,\; u_R$       & \multirow{4}{*}{$(\frac12,0) \;,\; (0,\frac12)$} & $\mathbf{3}$ & $\frac23$
          & \multirow{4}{*}{$\frac32$} \\
$d_L \;,\; d_R$       &                                 & $\mathbf{3}$ & $-\frac13$     & \\
$\nu_L \;,\; (\nu_R)$ &                                 & $\mathbf{1}$ & $0$            & \\
$e_L \;,\; e_R$       &                                 & $\mathbf{1}$ & $-1$           & \\
\midrule
$G_L \;,\; G_R$         & \multirow{4}{*}{$(1,0) \;,\; (0,1)$} & $\mathbf{8}$ & $0$ & \multirow{4}{*}{$2$}  \\
$W_L^\pm \;,\; W_R^\pm$ &                                      & $\mathbf{1}$ & $\pm 1$ & \\
$Z_L \;,\; Z_R$         &                                      & $\mathbf{1}$ & $0$ & \\
$A_L \;,\; A_R$         &                                      & $\mathbf{1}$ & $0$ & \\
\midrule
$V^\pm$ & \multirow{2}{*}{$(\frac{1}{2},\frac{1}{2})$} & \multirow{2}{*}{$\mathbf{1}$} & $\pm 1$ & \multirow{2}{*}{$1$}  \\
$V^z$   &                                              &                               & $0$     & \\
\midrule
$h$ & $(0,0)$ & $\mathbf{1}$ & $0$ & $1$ \\
\bottomrule
\end{tabular}
\caption{($\nu$)HEFT field representations under spacetime and internal symmetry groups. The corresponding canonical mass dimensions are also listed. The linearized Goldstone fields $V^\pm, V^z$ have mass dimension one, as they contain one power of derivative.}
\label{tbl:FieldContent}
\end{table}

\subsection{Compute $\Hcal^h$---Treating Powers of $h$ as Independent}
\label{subsec:Hh}

Given the field content and their (canonical mass dimension) power counting in \cref{tbl:FieldContent}, one is ready to compute a Hilbert series
\begin{equation}
\Hcal^h \left( q, \cD, \{\Phi\}, n_u, n_d, n_e, n_\nu \right) \,,
\end{equation}
which is an analytic function at the origin of the arguments $q, \cD, \{\Phi\}$ (and hence has a Taylor series expansion). Once Taylor expanded at the origin, each term corresponds to an EFT operator. The argument $\{\Phi\}$ collectively denotes all the fields given in \cref{tbl:FieldContent} (and for complex fields, their hermitian conjugates as well). Therefore, the powers of $\{\Phi\}$ indicate the field structure of the EFT operator. The power of $\cD$ corresponds to the number of derivatives in the operator, and the power of $q$ gives the mass dimension of the operator. We are also accommodating a generic flavor number structure, with $n_u$ species of $(u_L, u_R)$, $n_d$ species of $(d_L, d_R)$, $n_e$ species of $(e_L, e_R)$, and $n_\nu$ species of $(\nu_L, \nu_R)$. When $\{\Phi\}$ includes all the fields listed in \cref{tbl:FieldContent}, the resulting Hilbert series is for $\nu$HEFT. One can then readily select out the part for HEFT by sending the grading variable for right-handed neutrino to zero:
\begin{equation}
\Hcal^h_\text{HEFT} = \Hcal^h_{\nu\text{HEFT}}\, (\nu_R=0, \nu_R^\dagger=0) \,.
\label{eqn:HEFTfromnuHEFT}
\end{equation}

As explained in \cref{sec:Recap}, specifically \cref{eqn:HFull}, the full Hilbert series consists of two parts:
\begin{equation}
\Hcal^h = \Hcal^h_0 + \Delta\Hcal^h \,.
\label{eqn:HhSplit}
\end{equation}
The $\Hcal^h_0$ piece can be worked out automatically following \cref{eqn:H0}. The key task in this part is to work with the correct SPMs and their characters, as emphasized around \cref{eqn:Zdecompose}. We have reviewed and discussed the relevant ingredients for this in \cref{sec:SPMs}; see the bullet points at the beginning of this section for a summary.

The $\Delta\Hcal^h$ piece needs to be determined by finding all the co-closed but not co-exact form operators as described around \cref{eqn:cohomology,eqn:HFull,eqn:DeltaH}. This process requires some manual enumeration. Because the contributions to $\Delta\Hcal$ from the form operators that do not involve Goldstone fields (the upper section in \cref{tbl:CohomologyForms}) can be automatically worked out using conformal representation theory (see Section 4.1 in \cite{Henning:2017fpj}), we split $\Delta\Hcal^h$ as
\begin{equation}
\Delta\Hcal^h = \Delta\Hcal^h_\text{nonGoldstone} + \Delta\Hcal^h_\text{Goldstone} \,.
\label{eqn:DeltaHhSplit}
\end{equation}
We compute the non-Goldstone piece using conformal representation theory and obtain
\begin{align}
\Delta\Hcal^h_\text{nonGoldstone} &= q^3 h \cD^2 
+ q^4 \Big[ -\cD^4 - \left( Z_L + Z_R + A_L + A_R \right) \cD^2
+ n_u^2 \left( u_L u_L^\dagger + u_R u_R^\dagger \right) \cD
\notag\\[5pt]
&\hspace{80pt}
+ n_d^2 \left( d_L d_L^\dagger + d_R d_R^\dagger \right) \cD
+ n_e^2 \left( e_L e_L^\dagger + e_R e_R^\dagger \right) \cD
\notag\\[5pt]
&\hspace{80pt}
+ n_\nu^2 \left( \nu_L \nu_L^\dagger + \nu_R \nu_R^\dagger + \nu_L \nu_R + \nu_L^\dagger \nu_R^\dagger \right) \cD \Big] \,,
\label{eqn:DeltaHhnonGoldstone}
\end{align}
which clearly agrees with \cref{tbl:CohomologyForms} (upon suppressing the derivative power grading variable $\cD$). For the Goldstone piece, we use \cref{tbl:CohomologyForms} and \cref{eqn:DeltaH} together and get
\begin{align}
\Delta\Hcal^h_\text{Goldstone} &= q^2 V^z \cD + q^4 \Big[ V^z\cD^3 - V^+V^-\cD^2 + V^+V^-V^z\cD + V^z \left( Z_L + Z_R + A_L + A_R \right) \cD
\notag\\[5pt]
&\hspace{80pt}
+ V^+ \left( W_L^- + W_R^- \right) \cD + V^- \left( W_L^+ + W_R^+ \right) \cD \Big] \,.
\label{eqn:DeltaHhGoldstone}
\end{align}

Now putting everything together according to \cref{eqn:HhSplit,eqn:DeltaHhSplit}, we obtain the full Hilbert series $\Hcal^h$, which we organize according to the canonical mass dimensions
\begin{equation}
\Hcal^h = \sum_{\text{dim}=0}^\infty q^\text{dim}\, \Hcal^h_\text{dim} \,.
\label{eqn:HhOrders}
\end{equation}
The $\Hcal^h_\text{dim}$ at each mass dimension order is typically lengthy due to having a variety of fields, as well as a generic flavor number structure $(n_u, n_d, n_e, n_\nu)$. The full result with all of these information up to $\dim=7$ is contained in our ancillary file.

In order to make a tractable presentation, we merge grading variables of the Hilbert series as following
\begin{subequations}\label{eqn:GradingMerge}
\begin{align}
u_L \;,\; u_R \;,\; d_L \;,\; d_R
&\quad\longrightarrow\quad Q \,, \\[5pt]
u_L^\dagger \;,\; u_R^\dagger \;,\; d_L^\dagger \;,\; d_R^\dagger
&\quad\longrightarrow\quad \Qbar \,, \\[5pt]
e_L \;,\; e_R \;,\; \nu_L \;,\; \nu_R
&\quad\longrightarrow\quad L \,, \\[5pt]
e_L^\dagger \;,\; e_R^\dagger \;,\; \nu_L^\dagger \;,\; \nu_R^\dagger
&\quad\longrightarrow\quad \Lbar \,, \\[5pt]
G_L \;,\; G_R \;,\; W^\pm_L \;,\; W^\pm_R \;,\; Z_L \;,\; Z_R \;,\; A_L \;,\; A_R
&\quad\longrightarrow\quad X \,, \\[5pt]
V\pm \;,\; V_z
&\quad\longrightarrow\quad V \,,
\end{align}
\end{subequations}
and also take universal flavor numbers
\begin{equation}
n_u=n_d=n_e=n_\nu=n_f \,.
\label{eqn:universalnf}
\end{equation}
With these merged grading schemes, we could list the first few orders of $\Hcal^h$ in \cref{eqn:HhOrders}:
\begin{subequations}\label{eqn:Hhdemo}
\begin{align}
\Hcal^h_0 &= 1 \,,\\[6pt]
\Hcal^h_1 &= h \,,\\[6pt]
\Hcal^h_2 &= h^2 + 2\, V^2 \,,\\[6pt]
\Hcal^h_3 &= h^3 + 2\, h V^2 + 4n_f^2\, (L \Lbar + Q \Qbar) + (n_f^2 + n_f)\, (L^2 + \Lbar^2) \,,\\[6pt]
\Hcal^h_4 &= h^4 + 2\, h^2 V^2 + 4n_f^2\, h (L \Lbar + Q \Qbar) + (n_f^2 + n_f)\, h (L^2 + \Lbar^2)
\notag\\[3pt]
&\hspace{20pt}
+ 5\, V^4 + 8\, V^2 X + 10\, X^2 
+ 8n_f^2\, (L \Lbar + Q \Qbar) V 
+ 3n_f^2\, (L^2 + \Lbar^2) V \,,\\[6pt]
\Hcal^h_5 &= h^5 + 2\, h^3 V^2 + 4n_f^2\, h^2 (L \Lbar + Q \Qbar) + (n_f^2 + n_f)\, h^2 (L^2 + \Lbar^2)
\notag\\[3pt]
&\hspace{20pt}
+ 5\, h V^4 + 8\, h V^2 X + 10\, h X^2 + 8n_f^2\, h (L \Lbar + Q \Qbar) V + 3n_f^2\, h (L^2 + \Lbar^2) V
\notag\\[3pt]
&\hspace{20pt}
+ 4\, h V^3 \cD + 20n_f^2\, (L \Lbar + Q \Qbar) V^2 + 12n_f^2\, L \Lbar X + 16n_f^2\, Q \Qbar X
\notag\\[3pt]
&\hspace{20pt}
+ (8n_f^2 + 2n_f)\, (L^2 + \Lbar^2) V^2 + (4n_f^2 - 2n_f)\, (L^2 + \Lbar^2) X \,.
\end{align}
\end{subequations}
Higher order results $\Hcal^h_{\dim\ge6}$ are still lengthy even with the above merged grading, so we will not show them here.

\subsection{Compute $\Hcal^\text{FF}$---Wrapping Powers of $h$ into Form Factors}
\label{subsec:HFF}

As HEFT was historically introduced as an electroweak chiral Lagrangian without a physical Higgs field $h$, it is nowadays customary to wrap powers of the field $h$ into form factors, as opposed to counting them as independent operators. For example, this is saying that for vector bosons and fermions, we will view their mass terms $V^2, L\Lbar, Q\Qbar$ (contained in $\Hcal_2^h$ and $\Hcal_3^h$ in \cref{eqn:Hhdemo}) and their various interactions with the physical Higgs field $h$ (contained in higher $\Hcal_{\dim}^h$ in \cref{eqn:Hhdemo}) as one effective operator:
\begin{subequations}\label{eqn:Wrapping}
\begin{align}
V^2 \;,\; h V^2 \;,\; h^2 V^2 \;,\; \cdots &\quad\longrightarrow\quad \mathcal{F}(h)\, V^2 \,, \\[5pt]
L\Lbar \;,\; h L\Lbar \;,\; h^2 L\Lbar \;,\; \cdots &\quad\longrightarrow\quad \mathcal{Y}_L(h)\, L\Lbar \,, \\[5pt]
Q\Qbar \;,\; h Q\Qbar \;,\; h^2 Q\Qbar \;,\; \cdots &\quad\longrightarrow\quad \mathcal{Y}_Q(h)\, Q\Qbar \,.
\end{align}
\end{subequations}
In each line above, the various operators certainly all have distinct phenomenological consequences, and hence are not the ``same'' operator in the usual sense of an EFT. However, if one insists on the wrapping in \cref{eqn:Wrapping} and treating them as a single operator (for whatever reason), one is then introducing a new equivalence (redundancy) relation among the operators in $\Hcal^h$:
\begin{equation}
\cO_1 \sim \cO_2 \,,
\quad\text{if}\quad
\cO_1 = f(h)\, \cO_2
\quad\text{or}\quad
\cO_2 = f(h)\, \cO_1 \,.
\end{equation}
In \cref{sec:Spurion}, we explained how to systematically handle this new type of redundancy relation. Let $\Hcal^\text{FF}$ denote the new Hilbert series after wrapping into Form Factors; it can be simply obtained from $\Hcal^h$ as
\begin{equation}
\Hcal^\text{FF} = (1 - q h)\, \Hcal^h  \,.
\label{eqn:HFF}
\end{equation}

\begin{table}[t]
\renewcommand{\arraystretch}{1.5}
\setlength{\tabcolsep}{1.0em}
\centering\small
\begin{tabular}{cccc}
\toprule
Operator Class & Restrictions & HEFT & $\nu$HEFT \\
\midrule
$D^4$        &                & $15$ & $15$ \\
$D^2 X$      &                & $ 8$ & $ 8$ \\
$X^2$        &                & $10$ & $10$ \\
$X^3$        &                & $ 6$ & $ 6$ \\
\midrule\midrule
$\psi^2 D^2$ &                & $76 n_f^2 + 4 n_f$ & $110 n_f^2 + 6 n_f$ \\
$\psi^2 D$   &                & $15 n_f^2$ & $22 n_f^2$ \\
$\psi^2 X$   &                & $26 n_f^2 - 2 n_f$ & $36 n_f^2 - 4 n_f$ \\
$\psi^4$     &                & $\frac14 n_f^2 (445 n_f^2 - 6 n_f + 25)$ & $\frac12 n_f^2 ( 385 n_f^2 + 6 n_f + 13)$ \\
\midrule
$\psi^2 D^2$ & \multirow{4}{*}{$B-L$} & $60 n_f^2$ & $80 n_f^2$ \\
$\psi^2 D$   &                & $13 n_f^2$ & $16 n_f^2$ \\
$\psi^2 X$   &                & $22 n_f^2$ & $28 n_f^2$ \\
$\psi^4$     &                & $\frac14 n_f^2 (335 n_f^2 - 6 n_f + 31)$ & $n_f^2 ( 125 n_f^2 - 2 n_f + 9 )$ \\
\midrule
$\psi^2 D^2$ & \multirow{4}{*}{$B$ and $L$} & $60 n_f^2$ & $80 n_f^2$ \\
$\psi^2 D$   &                & $13 n_f^2$ & $16 n_f^2$ \\
$\psi^2 X$   &                & $22 n_f^2$ & $28 n_f^2$ \\
$\psi^4$     &                & $\frac14 n_f^2 (275 n_f^2 + 6 n_f + 31)$ & $n_f^2 ( 105 n_f^2 + 2 n_f + 9 )$ \\
\bottomrule
\end{tabular}
\caption{Eight classes of operators that are relevant for ``NLO'' ($\nu$)HEFT. For the fermionic classes, we are presenting three different scenarios---(1) no additional Baryon number $B$ or Lepton number $L$ restrictions, (2) requiring $B-L$ conservation, and (3) requiring both $B$ and $L$ conservation.}
\label{tbl:OperatorClasses}
\end{table}

\begin{table}[t]
\renewcommand{\arraystretch}{1.35}
\setlength{\tabcolsep}{1.0em}
\centering\small
\begin{tabular}{ccccc}
\toprule
Class & Detailed Class  & dim & HEFT & $\nu$HEFT \\ \midrule
\multirow{5}{*}{$D^4$} & $V^4$           & 4 & \multicolumn{2}{c}{$5$} \\
                       & $h \cD V^3$     & 5 & \multicolumn{2}{c}{$4$} \\
                       & $h^2 \cD^2 V^2$ & 6 & \multicolumn{2}{c}{$4$} \\
                       & $h^3 \cD^3 V$   & 7 & \multicolumn{2}{c}{$1$} \\
                       & $h^4 \cD^4$     & 8 & \multicolumn{2}{c}{$1$} \\
\midrule
$D^2 X$ & $V^2 X$      & 4 & \multicolumn{2}{c}{$ 8$} \\
\midrule
$X^2$ &                & 4 & \multicolumn{2}{c}{$10$} \\
\midrule
$X^3$ &                & 6 & \multicolumn{2}{c}{$ 6$} \\
\midrule
\multirow{3}{*}{$\psi^2 D^2$} & $L\Lbar V^2 \;,\; Q\Qbar V^2$             & 5 & $10 n_f^2 \;,\; 20 n_f^2$ & $20 n_f^2 \;,\; 20 n_f^2$ \\
                              & $L\Lbar h \cD V \;,\; Q\Qbar h \cD V$     & 6 & $ 8 n_f^2 \;,\; 16 n_f^2$ & $16 n_f^2 \;,\; 16 n_f^2$ \\
                              & $L\Lbar h^2 \cD^2 \;,\; Q\Qbar h^2 \cD^2$ & 7 & $ 2 n_f^2 \;,\;  4 n_f^2$ & $4 n_f^2 \;,\;  4 n_f^2$ \\
\midrule
$\psi^2 D$ & $L\Lbar V \;,\; Q\Qbar V$ & 4 & $5 n_f^2 \;,\;  8 n_f^2$ & $ 8 n_f^2 \;,\;  8 n_f^2$ \\
\midrule
$\psi^2 X$ & $L\Lbar X \;,\; Q\Qbar X$ & 5 & $6 n_f^2 \;,\; 16 n_f^2$ & $12 n_f^2 \;,\; 16 n_f^2$ \\
\midrule
\multirow{3}{*}{$\psi^4$} & $(L\Lbar)^2$       & 6 & $\frac14 n_f^2 (19 n_f^2 + 6 n_f + 7)$ & $n_f^2 (15 n_f^2 + 2 n_f + 3)$ \\
                          & $(L\Lbar)(Q\Qbar)$ & 6 & $34 n_f^4$            & $60 n_f^4$ \\
                          & $(Q\Qbar)^2$       & 6 & $n_f^2 (30n_f^2 + 6)$ & $n_f^2 (30 n_f^2 + 6)$ \\
\midrule\midrule
\multirow{3}{*}{$\psi^2 D^2$} & $L^2 V^2 + \hc$       & 5 & $\frac12 (9 n_f^2 + 3 n_f)$ & $8 n_f^2 + 2 n_f$ \\
                              & $L^2 h \cD V + \hc$   & 6 & $3 n_f^2$                   & $6 n_f^2$ \\
                              & $L^2 h^2 \cD^2 + \hc$ & 7 & $\frac12 (n_f^2 + n_f)$     & $n_f^2 + n_f$ \\
\midrule
$\psi^2 D$ & $L^2 V + \hc$         & 4 & $n_f^2$ & $3 n_f^2$ \\
\midrule
$\psi^2 X$ & $L^2 X + \hc$         & 5 & $2 n_f^2 - n_f$ & $4 n_f^2 - 2 n_f$ \\
\midrule
\multirow{5}{*}{$\psi^4$} & $L^4 + \hc$       & 6 & $\frac{1}{12} n_f^2 (n_f^2 - 1)$       & $\frac{1}{12} n_f^2 (5 n_f^2 + 6 n_f + 1)$ \\
                          & $L^3\Lbar + \hc$  & 6 & $\frac12 n_f^2 (3 n_f^2 + n_f)$        & $\frac13 n_f^2 (20 n_f^2 + 6 n_f - 2)$ \\
                          & $L^2Q\Qbar + \hc$ & 6 & $n_f^2 (8 n_f^2 + n_f)$                & $n_f^2 (20 n_f^2 + 2 n_f)$ \\
                          & $L\Qbar^3 + \hc$  & 6 & $\frac16 n_f^2 (25 n_f^2 - 9 n_f - 4)$ & $\frac13 n_f^2 (20 n_f^2 - 6 n_f - 2)$ \\
                          & $LQ^3 + \hc$      & 6 & $\frac12 n_f^2 (15 n_f^2 - 3 n_f)$     & $n_f^2 (10 n_f^2 - 2 n_f)$ \\
\bottomrule
\end{tabular}
\caption{Detailed breakdown of the eight classes of operators that are relevant for ``NLO'' ($\nu$)HEFT. The upper section contains operators preserving both the Baryon number $B$ and the Lepton number $L$, while the lower section contains operators breaking $B$ and/or $L$.}
\label{tbl:DetailedClasses}
\end{table}

Again, we organize $\Hcal^\text{FF}$ according to the canonical mass dimension
\begin{equation}
\Hcal^\text{FF} = \sum_{\text{dim}=0}^\infty q^\text{dim}\, \Hcal^\text{FF}_\text{dim} \,.
\label{eqn:HFFOrders}
\end{equation}
Using the same merged grading variables in \cref{eqn:GradingMerge} and the universal flavor number $n_f$ in \cref{eqn:universalnf}, its first few orders are
\begin{subequations}\label{eqn:HFFdemo}
\begin{align}
\Hcal^\text{FF}_0 &= 1 \,,\\[6pt]
\Hcal^\text{FF}_1 &= 0 \,,\\[6pt]
\Hcal^\text{FF}_2 &= 2\, V^2 \,,\\[6pt]
\Hcal^\text{FF}_3 &= 4n_f^2\, (L \Lbar + Q \Qbar) + (n_f^2 + n_f)\, (L^2 + \Lbar^2) \,,\\[6pt]
\Hcal^\text{FF}_4 &= 5\, V^4 + 8\, V^2 X + 10\, X^2 + 8n_f^2\, (L \Lbar + Q \Qbar) V + 3n_f^2\, (L^2 + \Lbar^2) V \,,\\[6pt]
\Hcal^\text{FF}_5 &= 4\, h V^3 \cD + 20n_f^2\, (L \Lbar + Q \Qbar) V^2 + 12n_f^2\, L \Lbar X + 16n_f^2\, Q \Qbar X
\notag\\[3pt]
&\hspace{20pt}
+ (8n_f^2 + 2n_f)\, (L^2 + \Lbar^2) V^2 + (4n_f^2 - 2n_f)\, (L^2 + \Lbar^2) X \,.
\end{align}
\end{subequations}
which we can readily obtain from \cref{eqn:Hhdemo} using \cref{eqn:HFF}. Similar with $\Hcal^h$, higher dimension results of $\Hcal^\text{FF}$ get lengthy even with the above merged grading scheme. These results are available in our ancillary file, but are not very enlightening to present here. Instead, let us focus on the following eight classes of operators
\begin{equation}
D^4 \;,\; D^2 X \;,\; X^2 \;,\; X^3 \;,\; \psi^2 D^2 \;,\; \psi^2 D \;,\; \psi^2 X \;,\; \psi^4 \,,
\label{eqn:eightclasses}
\end{equation}
which are typically discussed in the context of the ``NLO operators'' for ($\nu$)HEFT in the literature. Here notation is further condensed. Each ``$D$'' represents a derivative acting on the scalar field, which could be either the linearized Goldstone field $V$ or the combination $h\cD$; each fermion ``$\psi$'' can be either a lepton or quark or their conjugate, namely
\begin{subequations}
\begin{align}\label{eqn:Dpsi}
D &\in \{ V \;,\; h \cD \} \,, \\[5pt]
\psi &\in \{ L \;,\; \Lbar \;,\; Q \;,\; \Qbar \} \,.
\end{align}
\end{subequations}
We see that each ``$D$'' could have canonical dimension one or two, so for finding these eight classes of operators, it is sufficient to compute $\Hcal^\text{FF}$ up to canonical mass dimension eight. In \cref{tbl:OperatorClasses}, we summarize the counting results of these eight classes of operators. For the fermionic classes, we have included three different scenarios---(1) no additional Baryon number $B$ or Lepton number $L$ restrictions, (2) requiring $B-L$ conservation, and (3) requiring both $B$ and $L$ conservation. A more detailed breakdown of the counting of these eight classes of operators is provided in \cref{tbl:DetailedClasses}. Our results agree with the available ones in Ref. \cite{Brivio:2016fzo,Buchalla:2013rka,Krause:2016uhw}, up to the errors that are already pointed out by Ref.~\cite{Sun:2022ssa}. However, we note that the ``all operators'' result claimed in \cite{Sun:2022ssa} actually agree with our scenario of imposing $B-L$ symmetry in the $\nu$HEFT case. We still disagree with this reference on the result for the four-fermion class in the HEFT case, even with the $B-L$ conservation imposed.

\subsection{($\nu$)HEFT Hilbert Series by Spurion Approach}\label{subsec:HEFT_hs_spurion}

We know that the electroweak symmetry breaking structure is $SU(2)_L \times U(1)_Y \to U(1)_\text{EM}$. However, when it comes to a nonlinear realization, it is often preferred to consider a ``custodial upgrade'' of it, which is
\begin{equation}
SU(2)_L \times SU(2)_R \quad\longrightarrow\quad SU(2)_V \,,
\label{eqn:CustodialSB}
\end{equation}
where $SU(2)_V$ denotes the diagonal subgroup of the two $SU(2)$'s on the left-hand side. This upgraded version also has three broken generators and yields three Goldstones that would generate masses for the electroweak gauge bosons via Higgs mechanism.

Although one could argue for phenomenological incentives, the main motivation for considering the upgrade in \cref{eqn:CustodialSB} is a technical one. The symmetry breaking structure in \cref{eqn:CustodialSB} is a symmetric one (\textit{i.e.} \(G/H\) is a symmetric space), same as the two-flavor version of the QCD chiral Lagrangian. In this case, it is well known~\cite{Coleman:1969sm} that one could save the effort of using the Cartan-Maurer linearization variable and instead directly use the squared Goldstone matrix $U\equiv\xi^2$ as the linearly transforming building block. This is because under the symmetry breaking structure in \cref{eqn:CustodialSB}, the Goldstone matrix $\xi$ transforms as
\begin{equation}
\xi \quad\longrightarrow\quad \tilde\xi = g_L\, \xi\, h^{-1} = h\, \xi\, g_R^\dagger \,\,,
\end{equation}
where $h$ is a nonlinear compensating matrix (not to be confused with the physical Higgs field). We see that one could square the Goldstone matrix $\xi$ to cancel the nonlinear transforming factor $h$:
\begin{equation}
U \equiv \xi^2 \qquad\Longrightarrow\qquad
U \;\;\longrightarrow\;\; \tilde{U} = g_L\, U\, g_R^\dagger \,\,.
\end{equation}
Therefore, a linearly transforming building block $U$ is readily obtained.

\begin{table}[t]
\renewcommand{\arraystretch}{1.3}
\setlength{\arrayrulewidth}{.2mm}
\setlength{\tabcolsep}{1em}
\centering
\begin{tabular}{cccccc}
\toprule
Field     & Lorentz Group                   & $SU(3)_C$                     & $SU(2)_V$ & $\text{dim}$ \\
\midrule
$Q_L \;,\; Q_R$ & \multirow{2}{*}{$(\frac12,0) \;,\; (0,\frac12)$} & $\mathbf{3}$ & $\mathbf{2}$
                & \multirow{2}{*}{$\frac32$} \\
$L_L \;,\; L_R$ &                                                  & $\mathbf{1}$ & $\mathbf{2}$ & \\
\midrule
$G_L \;,\; G_R$ & \multirow{3}{*}{$(1,0) \;,\; (0,1)$} & $\mathbf{8}$ & $\mathbf{1}$ & \multirow{3}{*}{$2$}  \\
$W_L \;,\; W_R$ &                                      & $\mathbf{1}$ & $\mathbf{3}$ & \\
$B_L \;,\; B_R$ &                                      & $\mathbf{1}$ & $\mathbf{1}$ & \\
\midrule
$V$             & $(\frac{1}{2},\frac{1}{2})$ & $\mathbf{1}$ & $\mathbf{3}$ & $1$ \\
\midrule
$h$             & $(0,0)$                     & $\mathbf{1}$ & $\mathbf{1}$ & $1$ \\
\midrule
$T$             & $(0,0)$                     & $\mathbf{1}$ & $\mathbf{3}$ & $0$ \\
\bottomrule
\end{tabular}
\caption{($\nu$)HEFT field representations under spacetime and internal symmetry groups when the $U(1)_\text{EM}$ is promoted to the custodial symmetry $SU(2)_V$. We have included a spurion field $T$, which is an $SU(2)_V$ adjoint that would break it down to its Cartan subgroup $U(1)_V$.}
\label{tbl:FieldContentSpu}
\end{table}

Saving the CCWZ linearization effort is great. However, it does not come for free. After all, the SM does not preserve the custodial symmetry $SU(2)_V$; only $U(1)_\text{EM}$ is the true symmetry. In particular, there are two sources of $SU(2)_V$ breaking that need to be allowed in SM (and hence ($\nu$)HEFT). One is hidden in the covariant derivative $D_\mu$---the lack of gauge field components to form a full $SU(2)_R$ adjoint representation. To handle this issue, one typically chooses to work with the matrix field
\begin{equation}
V_\mu \equiv \left(D_\mu U\right) U^\dagger \,,
\label{eqn:Vdef}
\end{equation}
as opposed to using the matrix $U$ itself. $V_\mu$ forms an $SU(2)_L$ adjoint, and hence an $SU(2)_V$ adjoint. The other source of $SU(2)_V$ breaking allowed in ($\nu$)HEFT is that only its $U(1)_V$ subgroup generated by the $t_V^3$ generator is required to be preserved. This source can be handled by introducing a spurion field
\begin{equation}
T \equiv U \sigma^3 U^\dagger \,,
\label{eqn:Tdef}
\end{equation}
which transforms as an $SU(2)_L$ adjoint, and hence an $SU(2)_V$ adjoint. In such a spurion approach, other SM fields are also organized into custodial $SU(2)_V$ representations. A summary of the field representations is given in \cref{tbl:FieldContentSpu}.

With the field content and their power counting in \cref{tbl:FieldContentSpu}, we are now ready to compute the Hilbert series
\begin{equation}
\Hcal^h_\text{C} \left( q, \cD, \{\Phi_\text{C}\}, T, n_Q, n_L \right) \,,
\end{equation}
where we are using a subscript ``C'' to denote the custodial upgrade. The argument $\{\Phi_\text{C}\}$ collectively denotes all the dynamic fields given in \cref{tbl:FieldContentSpu}, with the spurion field $T$ separated out. A generic flavor number structure is now described by $n_Q, n_L$.

One might wonder how to select out the HEFT part of the $\nu$HEFT Hilbert series, given that in this custodial upgrade, $\nu_R$ and $e_R$ are now grouped into the $SU(2)_V$ doublet $L_R$. This can be achieved by taking a finer grading of the $L_R$ character. Specifically, one alters the internal symmetry parts of the $L_R, L_R^\dagger$ SPM characters as
\begin{subequations}\label{eqn:FinerGrading}
\begin{align}
L_R\, \chi_{L_R}^{} \propto L_R\, \chi^{SU(2)}_\mathbf{2}(w) = L_R \left( w + w^{-1} \right) 
&\quad\longrightarrow\quad
\nu_R\, w + e_R\, w^{-1} \,,\\[5pt]
L_R^\dagger\, \chi_{L_R^\dagger}^{} \propto L_R^\dagger\, \chi^{SU(2)}_\mathbf{2}(w) = L_R^\dagger \left( w + w^{-1} \right) 
&\quad\longrightarrow\quad
e_R^\dagger\, w + \nu_R^\dagger\, w^{-1} \,.
\end{align}
\end{subequations}
With the above finer grading, one could select out the HEFT part by taking $\nu_R=\nu_R^\dagger=0$ just as before (c.f. \cref{eqn:HEFTfromnuHEFT})
\begin{equation}
\Hcal^h_\text{C,\,HEFT} = \Hcal^h_{\text{C},\,\nu\text{HEFT}}\, (\nu_R=0, \nu_R^\dagger=0) \,.
\end{equation}

\begin{table}[t]
\renewcommand{\arraystretch}{1.2}
\setlength{\tabcolsep}{0.8em}
\centering
\begin{tabular}{cccc}
\toprule
Class & Detailed Class  & dim & ($\nu$)HEFT \\ \midrule
\multirow{5}{*}{$D^4$} & $V^4$           & 4 & $2+2T^2+T^4$ \\
                       & $h \cD V^3$     & 5 & $2T+T^2+T^3$ \\
                       & $h^2 \cD^2 V^2$ & 6 & $2+2T^2$ \\
                       & $h^3 \cD^3 V$   & 7 & $T$ \\
                       & $h^4 \cD^4$     & 8 & $1$ \\
\midrule
$D^2 X$ & $V^2 X$      & 4 & $2+4T+2T^2$ \\
\midrule
$X^2$ &                & 4 & $6+2T+2T^2$ \\
\midrule
$X^3$ &                & 6 & $4+2T$ \\
\bottomrule
\end{tabular}
\caption{Spurion structure of the detailed breakdown of the eight classes of operators that are relevant for ``NLO'' ($\nu$)HEFT, bosonic sector.}
\label{tbl:DetailedClassesSpuBosonic}
\end{table}

As explained earlier, $\Hcal^h_\text{C}$ has two pieces according to \cref{eqn:HFull}. Among these, only the part of $\Delta\Hcal^h_\text{C}$ that involves the Goldstone fields cannot be computed automatically. For this part, we use \cref{tbl:CohomologyForms} and \cref{eqn:DeltaH} to compute it manually. Applied to the field content in \cref{tbl:FieldContentSpu}, it leads us to
\begin{align}
\Delta\Hcal^h_\text{C,\,Goldstone} &= q^2\, V\, T + q^4 \bigg\{ V^3 + V (W_L + W_R) + V (W_L + W_R)\, T^2
\notag\\[5pt]
&\hspace{80pt}
+ \Big[ V - V^2 + V (B_L + B_R) + V (W_L + W_R) \Big] T \bigg\} \,.
\label{eqn:DeltaHTGoldstone}
\end{align}
Note that as an $SU(2)_V$ adjoint representation, the spurion field $T$ generates a trivial set of $SU(2)_V$ singlets
\begin{equation}
\Hcal^h_\text{C} (T, \{\Phi_\text{C}\} = 0) = \frac{1}{1-T^2} = 1 + T^2 + T^4 + T^6 + \cdots \,.
\label{eqn:HT}
\end{equation}
In \cref{eqn:DeltaHTGoldstone}, we are only keeping the nontrivial terms. For the other parts of $\Hcal^h_\text{C}$ that can be computed automatically using conformal representation theory, we also need to mod out the trivial set given in \cref{eqn:HT}. As explained in \cref{sec:Spurion}, this can be done by simply multiplying the factor
\begin{equation}
\frac{1}{\Hcal^h_\text{C} (T, \{\Phi_\text{C}\} = 0)} = 1 - T^2 \,.
\end{equation}

Taking the above treatment to remove the trivial singlet factors from the spurion $T$, and putting all parts together, we obtain the full $\Hcal^h_\text{C}$ organized in canonical mass dimension
\begin{equation}
\Hcal^h_\text{C} = \sum_{\dim=0}^\infty q^{\dim}\, \Hcal^h_\text{C,\,dim} \,.
\end{equation}
The detailed expression of $\Hcal^h_\text{C}$ up to $\dim=7$ is in our ancillary file. Here we emphasize that with the spurion introduced in \cref{eqn:Tdef}, one breaks the custodial symmetry $SU(2)_V$ to its Cartan $U(1)_V$ subgroup generated by the $t_V^3$ generator. However, note that this $U(1)_V$ is not the same as $U(1)_\text{EM}$, because
\begin{equation}
t_V^3 = t_L^3 + t_R^3 = t_L^3 + Y - \frac12 (B-L) = Q - \frac12 (B-L) \,.
\end{equation}
We see that $t_V^3 \ne Q$, and preserving $t_V^3$ is only equivalent to preserving $Q$ for the $B-L$ preserving sector, such as the Higgs sector. Therefore, the Hilbert series $\Hcal^h_\text{C}$ is not supposed to fully reproduce $\Hcal^h$ upon taking the $T\to 1$ limit (as well as unifying field grading variables). They are only supposed to agree for the $B-L$ preserving part. We have explicitly verified this:
\begin{subequations}
\begin{align}
\lim_{T\to 1} \Hcal^h_\text{C} &\ne \Hcal^h \,,\\[6pt]
\lim_{T\to 1} \Hcal^h_\text{C} \Big|_{B-L=0} &= \Hcal^h \Big|_{B-L=0} \,.
\end{align}
\end{subequations}

\begin{table}[t]
\renewcommand{\arraystretch}{1.35}
\setlength{\tabcolsep}{1.0em}
\centering\small
\begin{tabular}{ccccc}
\toprule
\multicolumn{2}{c}{Class} & dim & HEFT & $\nu$HEFT \\ \midrule
\multirow{6}{*}{$\psi^2 D^2$} & $L\Lbar V^2$       & 5 & $(2+4T+3T^2+T^3) n_f^2$  & $(4+8T+6T^2+2T^3) n_f^2$ \\
                              & $L\Lbar h \cD V$   & 6 & $(2+4T+2T^2) n_f^2$      & $(4+8T+4T^2) n_f^2$ \\
                              & $L\Lbar h^2 \cD^2$ & 7 & $(1+T) n_f^2$            & $(2+2T) n_f^2$ \\
                              & $Q\Qbar V^2$       & 5 & $(4+8T+6T^2+2T^3) n_f^2$ & $(4+8T+6T^2+2T^3) n_f^2$ \\
                              & $Q\Qbar h \cD V$   & 6 & $(4+8T+4T^2) n_f^2$      & $(4+8T+4T^2) n_f^2$ \\
                              & $Q\Qbar h^2 \cD^2$ & 7 & $(2+2T) n_f^2$           & $(2+2T) n_f^2$ \\
\midrule
\multirow{2}{*}{$\psi^2 D$} & $L\Lbar V$ & 4 & $(1+3T+ T^2) n_f^2$ & $(2+4T+2T^2) n_f^2$ \\
                            & $Q\Qbar V$ & 4 & $(2+4T+2T^2) n_f^2$ & $(2+4T+2T^2) n_f^2$ \\
\midrule
\multirow{2}{*}{$\psi^2 X$} & $L\Lbar X$ & 5 & $(2+3T+ T^2) n_f^2$ & $(4+6T+2T^2) n_f^2$ \\
                            & $Q\Qbar X$ & 5 & $(6+8T+2T^2) n_f^2$ & $(6+8T+2T^2) n_f^2$ \\
\midrule
         & \multirow{3}{*}{$(L\Lbar)^2$} & \multirow{3}{*}{6} & $\frac14 n_f^2 (7n_f^2+2n_f+7)$ & $n_f^2 (5n_f^2+3)$ \\
         &&& $+\frac14 n_f^2 (9n_f^2+2n_f-3) T$   & $+\frac12 n_f^2 (15n_f^2+2n_f-3) T$ \\
         &&& $+\frac14 n_f^2 (3n_f^2+2n_f+3) T^2$ & $+\frac12 n_f^2 ( 5n_f^2+2n_f+3) T^2$ \\
\cmidrule(){2-5}
$\psi^4$ & $(L\Lbar)(Q\Qbar)$ & 6 & $(12+17T+5T^2) n_f^4$ & $(20+30T+10T^2) n_f^4$ \\
\cmidrule(){2-5}
         & \multirow{3}{*}{$(Q\Qbar)^2$} & \multirow{3}{*}{6} & $n_f^2 (10n_f^2+6)$ & $n_f^2 (10n_f^2+6)$ \\
         &&& $+n_f^2 (15n_f^2-3) T$   & $+n_f^2 (15n_f^2-3) T$ \\
         &&& $+n_f^2 ( 5n_f^2+3) T^2$ & $+n_f^2 ( 5n_f^2+3) T^2$ \\
\bottomrule
\end{tabular}
\caption{Spurion structure of the detailed breakdown of the eight classes of operators that are relevant for ``NLO'' ($\nu$)HEFT, fermionic sector conserving $B$ and $L$.}
\label{tbl:DetailedClassesSpuFermionicBLcon}
\end{table}

The next step is to wrap powers of the physical Higgs field $h$ into form factors (similar to \cref{eqn:HFF})
\begin{equation}
\Hcal^\text{FF}_\text{C} = (1-qh)\, \Hcal^h_\text{C} = \sum_{\dim=0}^\infty q^{\dim}\, \Hcal^\text{FF}_\text{C,\,dim} \,.
\end{equation}
In order to present some example expressions of $\Hcal^\text{FF}_\text{C,\,dim}$, we adopt a merged grading scheme that is similar to \cref{eqn:GradingMerge}, but now for custodial upgraded version of the field variables:
\begin{subequations}
\begin{align}
Q_L \;,\; Q_R
&\quad\longrightarrow\quad Q \,, \\[5pt]
Q_L^\dagger \;,\; Q_R^\dagger
&\quad\longrightarrow\quad \Qbar \,, \\[5pt]
L_L \;,\; \nu_R \;,\; e_R 
&\quad\longrightarrow\quad L \,, \\[5pt]
L_L^\dagger \;,\; \nu_R^\dagger \;,\; e_R^\dagger 
&\quad\longrightarrow\quad \Lbar \,, \\[5pt]
G_L \;,\; G_R \;,\; W_L \;,\; W_R \;,\; B_L \;,\; B_R
&\quad\longrightarrow\quad X \,.
\end{align}
\end{subequations}
We recall from \cref{eqn:FinerGrading} that before performing the above grading merge, we actually had the finer grading $\nu_R, e_R, \nu_R^\dagger, e_R^\dagger$, instead of $L_R, L_R^\dagger$, such that one could readily select out the HEFT part of the Hilbert series. Let us also merge the flavor number structure into the universal one
\begin{equation}
n_Q = n_L = n_f \,.
\end{equation}
With this grading merge, the first few orders of $\Hcal^\text{FF}_\text{C}$ reads
\begin{subequations}
\begin{align}
\Hcal^\text{FF}_{\text{C},\,0} &= 1 \,,\\[8pt]
\Hcal^\text{FF}_{\text{C},\,1} &= 0 \,,\\[8pt]
\Hcal^\text{FF}_{\text{C},\,2} &= V^2\, (1 + T^2) \,,\\[8pt]
\Hcal^\text{FF}_{\text{C},\,3} &= (2+2T)\, n_f^2\, (L\Lbar + Q\Qbar) + \Big[ n_f^2-n_f + (n_f^2+n_f) T \Big] (L^2 + \Lbar^2) \,,\\[8pt]
\Hcal^\text{FF}_{\text{C},\,4} &= (2+2T^2+T^4)\, V^4 + (2+4T+2T^2)\, V^2 X + (6+2T+2T^2)\, X^2 
\notag\\[3pt]
&\hspace{20pt}
+ (2+4T+2T^2)\,n_f^2\, (L \Lbar + Q \Qbar) V + (1+2T+T^2)\,n_f^2\, (L^2 + \Lbar^2) V\,,\\[8pt]
\Hcal^\text{FF}_{\text{C},\,5} &= (2T+T^2+T^3)\, h V^3 \cD + (4+8T+6T^2+2T^3)\,n_f^2\, (L\Lbar + Q\Qbar) V^2
\notag\\[3pt]
&\hspace{20pt}
+ (4+6T+2T^2)\, n_f^2\, L\Lbar X 
+ (6+8T+2T^2)\, n_f^2\, Q\Qbar X
\notag\\[3pt]
&\hspace{20pt}
+ \Big[ (2+4T+3T^2+T^3)\, n_f^2 - (2-2T+T^2-T^3)\,n_f \Big]\, (L^2 + \Lbar^2) V^2
\notag\\[3pt]
&\hspace{20pt}
+ \Big[ (2+3T+T^2)\, n_f^2 - (T+T^2)\,n_f \Big]\, (L^2 + \Lbar^2) X
\,.
\end{align}
\end{subequations}
We see explicitly here that taking $T\to 1$ reproduces \cref{eqn:HFFdemo} only for the $B-L$ preserving sectors. Higher order results, which are too lengthy to be enlightening, are in our ancillary file. Instead, we focus again on the eight classes of operators in \cref{eqn:eightclasses}, and summarize their results in \cref{tbl:DetailedClassesSpuBosonic,tbl:DetailedClassesSpuFermionicBLcon,tbl:DetailedClassesSpuFermionicBLvio}. Comparing with \cref{tbl:DetailedClasses}, we see clearly the agreement/disagreement for the $B-L$ preserving/violating sectors.

\begin{table}[t]
\renewcommand{\arraystretch}{1.5}
\setlength{\tabcolsep}{0.8em}
\centering\small
\begin{tabular}{ccccc}
\toprule
\multicolumn{2}{c}{Class} & dim & ``HEFT'' & ``$\nu$HEFT'' \\ \midrule
\multirow{4}{*}{$\psi^2 D^2$} & \multirow{2}{*}{$L^2 V^2 + \hc$} & \multirow{2}{*}{5} & $\frac{2n_f^2-3n_f}{2} + \frac{5n_f^2+3n_f}{2}T$ & $2n_f^2 - 2n_f + (4n_f^2+2n_f)T$ \\
                              &&& $+\frac{7n_f^2-3n_f}{4}T^2+\frac{3n_f^2+3n_f}{4}T^3$ & $+(3n_f^2-n_f)T^2+(n_f^2+n_f)T^3$\\
\cmidrule(){2-5}
                              & $L^2 h \cD V + \hc$   & 6 & $\frac12(3+4T+3T^2)n_f^2$ & $(2+4T+2T^2)n_f^2$ \\
\cmidrule(){2-5}
                              & $L^2 h^2 \cD^2 + \hc$ & 7 & $\frac{n_f^2-3n_f}{4}+\frac{3n_f^2+3n_f}{4}T$ & $n_f^2-n_f+(n_f^2+n_f)T$ \\
\midrule
$\psi^2 D$ & $L^2 V + \hc$         & 4 & $\frac12(1+2T+T^2)n_f^2$ & $(1+2T+T^2)n_f^2$ \\
\midrule
\multirow{2}{*}{$\psi^2 X$} & \multirow{2}{*}{$L^2 X + \hc$} & \multirow{2}{*}{5} & $n_f^2+\frac14(7n_f^2-3n_f)T$ & $2n_f^2+(3n_f^2-n_f)T$ \\
                            &&& $+\frac14(3n_f^2-3n_f)T^2$ & $+(n_f^2-n_f)T^2$ \\
\midrule
                          & \multirow{3}{*}{$L^4 + \hc$} & \multirow{3}{*}{6} & $\frac{1}{12}n_f^2(2n_f^2+3n_f+13)$ & $\frac16n_f^2(5n_f^2+13)$ \\
                          &&& $+\frac13n_f^2(n_f^2-1)T$ & $+\frac14n_f^2(5n_f^2+2n_f-3)T$ \\
                          &&& $+\frac14n_f^2(n_f^2+n_f)T^2$ & $+\frac{1}{12}n_f^2(5n_f^2+6n_f+1)T^2$ \\
\cmidrule(){2-5}
                          & \multirow{3}{*}{$L^3\Lbar + \hc$} & \multirow{3}{*}{6} & $\frac{1}{12}n_f^2(11n_f^2+3n_f+4)$ & $\frac13n_f^2(10n_f^2+2)$ \\
                          &&& $+\frac12n_f^2(3n_f^2+n_f)T$ & $+n_f^2(5n_f^2+n_f)T$ \\
                          &&& $+\frac{1}{12}n_f^2(7n_f^2+3n_f-4)T^2$ & $+\frac13n_f^2(5n_f^2+3n_f-2)T^2$ \\
\cmidrule(){2-5}
\multirow{2}{*}{$\psi^4$} & \multirow{2}{*}{$L^2Q\Qbar + \hc$} & \multirow{2}{*}{6} & $5n_f^4+\frac14n_f^2(33n_f^2+3n_f)T$ & $10n_f^4+n_f^2(15n_f^2+n_f)T$ \\
                          &&& $+\frac14n_f^2(13n_f^2+3n_f)T^2$ & $+\frac13n_f^2(5n_f^2-3n_f-2)T^2$ \\
\cmidrule(){2-5}
                          & \multirow{3}{*}{$L\Qbar^3 + \hc$} & \multirow{3}{*}{6} & $\frac12n_f^2(5n_f^2+1)$ & $\frac13n_f^2(10n_f^2+2)$ \\
                          &&& $+\frac14n_f^2(15n_f^2-3n_f)T$ & $+n_f^2(5n_f^2-n_f)T$ \\
                          &&& $+\frac14n_f^2(5n_f^2-3n_f-2)T^2$ & $+\frac13n_f^2(5n_f^2-3n_f-2)T^2$ \\
\cmidrule(){2-5}
                          & \multirow{3}{*}{$LQ^3 + \hc$} & \multirow{3}{*}{6} & $\frac12n_f^2(5n_f^2+1)$ & $\frac13n_f^2(10n_f^2+2)$ \\
                          &&& $+\frac14n_f^2(15n_f^2-3n_f)T$ & $+n_f^2(5n_f^2-n_f)T$ \\
                          &&& $+\frac14n_f^2(5n_f^2-3n_f-2)T^2$ & $+\frac13n_f^2(5n_f^2-3n_f-2)T^2$ \\
\bottomrule
\end{tabular}
\caption{Spurion structure of the detailed breakdown of the eight classes of operators that are relevant for ``NLO'' ($\nu$)HEFT, fermionic sector violating $B$ and/or $L$. We emphasize that apart from the last row, $LQ^3 + \hc$, which preserves $B-L$, this table is not providing the correct result for the actual ($\nu$)HEFT, because for $B-L$ violating sectors, the custodial upgrade approach is not equivalent to the SM symmetry breaking structure; see text for detailed discussion. We include this table exactly for the purpose of comparing with the correct results in \cref{tbl:DetailedClasses} and demonstrating this disagreement.}
\label{tbl:DetailedClassesSpuFermionicBLvio}
\end{table}

\newpage
\section{Outlook}
\label{sec:Conclusions}

We expanded Hilbert series EFT methodologies to include massive particles in the spectrum of the theory. A number of interesting technical and conceptual results were obtained, in particular the manifestation of the Higgs mechanism in the Hilbert series and the analysis of spurion fields that take vevs. The application to HEFT was presented. We found agreement at NLO with original operator listings, and clarified issues of custodial symmetry spurion-based approaches, wherein the final operator basis obtained is inequivalent to a direct approach outside of the $B-L$ conserving sector. We conclude with some observations and avenues of potential interest for further study.

The way in which Hilbert series articulate the Higgs mechanism is expected from their interpretation as $S$-matrix partition functions. The elements and degeneracies of the $S$-matrix are the same viewed as either a massive vector or a massless vector plus a Nambu-Goldstone boson. The precise statement is that of the mode decomposition of the  massive single particle character; the generalization to $d$ spacetime dimensions and arbitrary spin $k$ particles points to applications of Hilbert series in massive gravity. 

Having showed that Hilbert series can easily accommodate the effects of spurion fields that take vevs, it is easy to envisage they could find additional useful application in phenomenological EFT (model) building, and analysis of symmetry breaking patterns. 

We look forward to investigations of the above, and to further understanding the structure of EFT through the lens of a Hilbert series.

\acknowledgments

We would like to thank Aneesh Manohar for useful discussions.
L.\,G.\ acknowledges support from the National Science Foundation, Grant PHY-1630782, and to the Heising-Simons Foundation, Grant 2017-228.
B.\,H.\ is supported by the Swiss National Science Foundation, under grants no. PP00P2-170578 and no. 200020-188671, and through the National Center of Competence in Research SwissMAP.
X.\,L.\ is supported by the U.S. Department of Energy, under grant number DE-SC0009919 and DE-SC0011640.
T.\,M.\ is supported by the World Premier International Research Center Initiative (WPI) MEXT, Japan, and by JSPS KAKENHI grants JP19H05810, JP20H01896, and JP20H00153.
The work of H.\,M.\ was supported by the Director, Office of Science, Office of High Energy Physics of the U.S. Department of Energy under the Contract No. DE-AC02-05CH11231, by the NSF grant PHY-1915314, by the JSPS Grant-in-Aid for Scientific Research JP20K03942, MEXT Grant-in-Aid for Transformative Research Areas (A) JP20H05850, JP20A203, by WPI, MEXT, Japan, the Institute for AI and Beyond of the University of Tokyo, and Hamamatsu Photonics, K.K.

------------------------------------------------------------------------------
\appendix
\newpage
\section{General Massive Single Particle Modules and the Higgs Mechanism}
\label{appsec:spm}

Let \(R_{\Ph}\) be the single particle module (SPM) for a particle/field \(\Ph\), which lives in spacetime dimension \(d\) with Euclidean Lorentz group \(SO(d)\). In this appendix we provide the mode decompositions for \(R_{\Ph}\),
\begin{equation}\label{eq:mode_decomp}
R_{\Ph} = \bigoplus_{l \in \L_{\Ph}}V_l \,,
\end{equation}
where \(V_l\) is a finite-dimensional irreducible representation (irrep) of \(SO(d)\) labeled by its highest weight vector \(l = (l_1,\dots,l_{\lfloor \frac{d}{2} \rfloor})\) and \(\L_{\Ph} = \{l\}\) is the set of \(SO(d)\) irreps appearing in the decomposition. We are searching to determine \(\L_{\Ph}\).

Once we have the mode decomposition~\eqref{eq:mode_decomp}, together with an understanding of the scaling dimension \(\D(l)\) of each \(V_l\) (this is easy---it's just counting derivatives on top of the field \(\Ph\)), the associated character \(\ch_{\Ph}^{}\) is given by
\begin{equation}
\ch^{}_{\Ph}(q,x) = \sum_{l\in \L_{\Ph}} q^{\D(l)}\ch_l(x) \,,
\end{equation}
where \(\ch^{}_l(x) \equiv \ch^{}_{(l_1,\dots,l_r)}(x_1,\dots,x_r)\), \(r = \lfloor \frac{d}{2} \rfloor\), is the \(SO(d)\) character for the representation \(V_l\). In many instances this sum can be performed directly, giving useful analytic expressions of the SPM character (e.g. Eq.~\eqref{eqn:A_char} for the character \(\chi^{}_A\) of a massive vector boson).

\subsection{Mode decomposition of massive SPMs from Frobenius reciprocity}

For a massive particle, \(p^2 = m^2\), the little group is \(SO(d-1)\), and the representation is built as an induced representation, \textit{i.e.} as a \(V_{\bar{l}}\)-valued function on the hyperboloid \(p^2 = m^2\), where \(V_{\bar{l}}\) is a  finite-dimensional irrep of \(SO(d-1)\).\footnote{The Euclidean version changes the hyperboloid to a sphere, i.e. \(SO(d)/SO(d-1) \simeq S^{d-1}\) while \(SO(d-1,1)/SO(d-1) \simeq H^{d-1}\). The two are related by analytic continuation. These differences don't matter for the computations in this appendix, so we loosely use Lorentzian and Euclidean language interchangeably.} The choice of \(V_{\bar{l}}\) physically corresponds to the spin of the particle. In summary, we are building
\begin{equation}
  R_{\Ph} = \text{Ind}_{SO(d-1)}^{SO(d)}(V_{\bar{l}_{\Ph}}) \,.
\end{equation}

One way to determine the mode decomposition of \(R_{\Ph}\) is via \textit{Frobenius reciprocity}, which gives an intuitive relation between induction and restriction. Let \(G\) be a compact, semi-simple Lie group, and \(H\subset G\) a subgroup. Let us denote irreps of \(G\) by \(V_l\) and those of \(H\) by \(V_{\bar{l}}\), where \(l = (l_1, \dots, l_{\text{rank}(G)})\) and \(\bar{l} = (\bar{l}_1, \dots, \bar{l}_{\text{rank}(H)})\) respectively denote the highest weight vectors of \(V_l\) and \(V_{\bar{l}}\). There are two basic operations we wish to consider---induction and restriction---which are dual to each other in a certain sense.

For induction, we ask: given \(V_{\bar{l}}\), what irreps of \(G\) show up in the induced representation? For restriction, we ask: given \(V_{l}\), how does this decompose into irreps of \(H\)? In other words, we want to determine the right hand side of the following equations:\footnote{\label{ft:ind_res_ex}Let's give examples of each, to make sure everyone is on the same page. Consider a function on the sphere \(S^{n-1} = SO(n)/SO(n-1)\). Such a function can be expanded in spherical harmonics, which transform irreducibly in the traceless symmetric representations of \(SO(n)\):
\begin{equation}
\text{Ind}_{SO(n-1)}^{SO(n)}(\mathbf{1}) = \bigoplus_{k=0}^{\infty}V_{(k,0,\dots,0)} \,. \nonumber
\end{equation}
Here, \((k,0,\dots,0)\) is the highest-weight vector for the rank-\(k\) traceless symmetric representation, which are often referred to as the ``spin-\(k\)'' representations of \(SO(n)\) in the physics literature. For restriction, consider the decomposition of the antisymmetric tensor of \(SO(n)\) into \(SO(n-1)\) irreps,
  \begin{equation}
  \begin{pmatrix}
    0 & A_{12} & A_{13} & \cdots & A_{1n} \\
    -A_{12} & 0 & A_{23} & \cdots & A_{2n} \\
    -A_{13} & -A_{23}  & \ddots & \cdots & \vdots \\
    \vdots & & & 0 & A_{n-1,n} \\
    -A_{1n} & -A_{2n} & \cdots & -A_{n-1,n}& 0
  \end{pmatrix} \to
  \begin{pmatrix}
    0 & \tcb{A_{12}} & \tcb{A_{13}} & \tcb{\cdots} & \tcb{A_{1n}} \\
    \tcb{-A_{12}} & \tcm{0} & \tcm{A_{23}} & \tcm{\cdots} & \tcm{A_{2n}} \\
    \tcb{-A_{13}} & \tcm{-A_{23}}  & \tcm{\ddots} & \tcm{\cdots} & \tcm{\vdots} \\
    \tcb{\vdots} & & & \tcm{0} & \tcm{A_{n-1,n}} \\
    \tcb{-A_{1n}} & \tcm{-A_{2n}} & \tcm{\cdots} & \tcm{-A_{n-1,n}}& \tcm{0}
  \end{pmatrix}, \nonumber
  \end{equation}
which obviously decomposes into a vector and anti-symmetric tensor of \(SO(n-1)\), as highlighted in blue and pink, respectively,
\begin{equation}
  \text{Res}^{SO(n)}_{SO(n-1)}(V_{(1,1,0,\dots,0)}) = \tcb{V_{(1,0,\dots,0)}} \oplus \tcm{V_{(1,1,0,\dots,0)}}\, . \nonumber
\end{equation}
}
\begin{align}
\text{Ind}_H^G(V_{\bar{l}}) &= \bigoplus V_l \,, \nonumber \\
\text{Res}_H^G(V_{l}) &= \bigoplus V_{\bar{l}} \,. \nonumber
\end{align}

Intuitively, if the restriction of the representation \(V_l\) contains \(V_{\bar{l}}\), \(\text{Res}^G_H(V_l) \supset V_{\bar{l}}\), then we might expect \(V_{l}\) to show up when we induce \(V_{\bar{l}}\). For example, the antisymmetric tensor \(\ytableausetup{boxsize=.5em,centertableaux} \ydiagram{1,1}\) of \(SO(n)\) contains the vector representation \(\ydiagram{1}\) of \(SO(n-1)\), so we likely anticipate that \(\text{Ind}_{SO(n-1)}^{SO(n)}(\ydiagram{1}) \supset \ydiagram{1,1}\). \textit{Frobenius reciprocity} makes this intuition precise: it says that the multiplicity for which the representation \(V_l\) shows up when inducing \(V_{\bar{l}}\) is equal to the multiplicity for which \(V_{\bar{l}}\) shows up when restricting \(V_l\),\footnote{Or, letting \(\langle \cdot,\cdot \rangle_G\) denote the \(G\)-invariant inner product,
\begin{equation}
\braket{\text{Ind}_H^G(V_{\bar{l}}),V_l}_G = \braket{V_{\bar{l}},\text{Res}^G_H(V_l)}_H. \nonumber
\end{equation}
}
\begin{equation}
\text{mult}\big(\text{Ind}_H^G(V_{\bar{l}}),V_l\big) = \text{mult}\big(V_{\bar{l}},\text{Res}^G_H(V_l)\big) \,.
\end{equation}
As an example to mull over, one can look over the spherical harmonics example given in footnote~\ref{ft:ind_res_ex}.

The application is hopefully obvious: if we know the restriction formulas for \(SO(d) \to SO(d-1)\), then we can reverse engineer them to determine the decomposition of the induced representation. Fortunately, these formulas are well known, \textit{e.g.} section 25.3 of~\cite{FultonHarris}. For the restriction of \(SO(2n+1) \to SO(2n)\) we have
\begin{equation}
\text{Res}_{SO(2n)}^{SO(2n+1)}(V_l) = \bigoplus V_{\bar{l}} \,, \nonumber
\end{equation}
where the sum is over all \(\bar{l} = (\bar{l}_1,\dots,\bar{l}_n)\) such that
\begin{equation}\label{eq:restrict_2r+1_to_2r}
l_1 \ge \bar{l}_1 \ge l_2 \ge \bar{l}_2 \ge \cdots \ge \bar{l}_{n-1} \ge l_n \ge \abs{\bar{l}_n} \,,
\end{equation}
with \(l_i-\bar{l}_i \in \mathbb{Z}\), \textit{i.e.} the \(l_i\) and \(\bar{l}_i\) are simultaneously either all integers or all half integers. For the restriction \(SO(2n) \to SO(2n-1)\) we have 
\begin{equation}
  \text{Res}_{SO(2n-1)}^{SO(2n)}(V_l) = \bigoplus V_{\bar{l}}, \nonumber
\end{equation}
where the sum is over all \(\bar{l} = (\bar{l}_1,\dots,\bar{l}_{n-1})\)  with
\begin{equation}\label{eq:restrict_2r_to_2r-1}
  l_1 \ge \bar{l}_1 \ge l_2 \ge \bar{l}_2 \ge \cdots \ge \bar{l}_{n-1} \ge \abs{l_n},
\end{equation}
with \(l_i-\bar{l}_i \in \mathbb{Z}\). As an illustration, consider the antisymmetric tensor from Footnote~\ref{ft:ind_res_ex}: we have \(l = (1,1,0,\dots,0)\), so that the restriction formula reads \(1 \ge \bar{l}_1 \ge 1 \ge \bar{l}_2 \ge 0\), which is solved by \(\bar{l} = (1,1,0,\dots,0)\) and \(\bar{l} = (1,0,\dots,0)\), i.e. \(\ytableausetup{boxsize=.5em,centertableaux} \text{Res}_{SO(n-1)}^{SO(n)} \big( \, \ydiagram{1,1} \, \big) = \ydiagram{1,1} \oplus \ydiagram{1}\).

Note that in the above restriction formulas for \(SO(n) \to SO(n-1)\) each \(V_{\bar{l}}\) shows up with unit multiplicity. Restricting \(SO(n)\) to \(SO(k)\) with \(k < n-1\) will generically lead to increased multiplicities, as can be determined by applying Eqs.~\eqref{eq:restrict_2r+1_to_2r} and~\eqref{eq:restrict_2r_to_2r-1} recursively, e.g. for \(k = n-2\), \(\ytableausetup{boxsize=.5em,centertableaux} \text{Res}_{SO(n-2)}^{SO(n)} \big( \, \ydiagram{1,1} \, \big) = \text{Res}_{SO(n-2)}^{SO(n-1)}\Big[\text{Res}_{SO(n-1)}^{SO(n)} \big( \, \ydiagram{1,1} \, \big) \Big] =  \text{Res}_{SO(n-2)}^{SO(n-1)} \big( \, \ydiagram{1,1} \oplus \ydiagram{1}\, \big)= \ydiagram{1,1} \oplus \ydiagram{1} \oplus \ydiagram{1} \oplus \mathbf{1}\). We note that this sort of analysis, together with the reverse engineering to determine the decomposition of the induced representation (see the examples below in Sec.~\ref{subsec:FrobeniusExamples}), explains the central result of~\cite{Henning:2019mcv}.\footnote{In particular, compare to Eq.~(30) in~\cite{Henning:2019mcv}. Although unrelated to the present discussion, reference~\cite{Henning:2019mcv} also highlights an interesting example where Frobenius reciprocity implies the conservation of higher spin currents in free theories.}

\subsection{Examples}\label{subsec:FrobeniusExamples}
\subsubsection{Massive spin-$k$}

As a first example consider massive spin-\(k\) particles, \textit{i.e.} those induced from \(V_{\bar{l}} = V_{(k,0,\dots,0)}\). The spin-0 (scalar) case corresponds to the spherical harmonics example in Footnote~\ref{ft:ind_res_ex}, which we leave to the reader to convince themselves (c.f. \cref{eqn:scalar_SPM}). For a massive spin-1 (vector) we have \((\bar{l}_1, \dots,\bar{l}_{\lfloor (d-1)/2\rfloor}) = (1,0,\dots,0)\) and the restriction formula reads
\begin{equation}
l_1 \ge 1 \ge l_2 \ge 0 \,, \nonumber
\end{equation}
which is solved by \((l_1,l_2) = (n+1,0)\) and \((n+1,1)\) for \(n=0,1,2,\dots\),
\begin{align}\label{eq:massive_spin-1}
\text{Spin-1:}~~\text{Ind}_{SO(d-1)}^{SO(d)}(\ydiagram{1}) =& \ydiagram{1} \oplus \ydiagram{2} \oplus \ydiagram{3} \oplus \ydiagram{4} \oplus \cdots \nonumber \\
  &\oplus  \ydiagram{1,1} \oplus \ydiagram{2,1} \oplus \ydiagram{3,1} \oplus \ydiagram{4,1} \oplus \cdots \,, \nonumber \\
  \Rightarrow~~\text{Ind}_{SO(d-1)}^{SO(d)}\big(V_{(1,0,\dots,0)}\big) =& \bigoplus_{n=0}^{\infty}\Big(V_{(n+1,0,\dots,0)} \oplus V_{(n+1,1,0,\dots,0)}\Big) \,.
\end{align}
As another example, a massive graviton obeys the restriction formula \(l_1 \ge 2 \ge l_2 \ge 0\):
\begin{align}\label{eq:massive_spin-2}
  \text{Spin-2:}~~\text{Ind}_{SO(d-1)}^{SO(d)}(\ydiagram{2}) =& \ydiagram{2} \oplus \ydiagram{3} \oplus \ydiagram{4} \oplus \ydiagram{5} \oplus \cdots \nonumber \\
  &\oplus  \ydiagram{2,1} \oplus \ydiagram{3,1} \oplus \ydiagram{4,1} \oplus \ydiagram{5,1} \oplus \cdots \,, \nonumber \\
  &\oplus  \ydiagram{2,2} \oplus \ydiagram{3,2} \oplus \ydiagram{4,2} \oplus \ydiagram{5,2} \oplus \cdots \,, \nonumber \\
  \Rightarrow~~\text{Ind}_{SO(d-1)}^{SO(d)}\big(V_{(2,0,\dots,0)}\big) =& \bigoplus_{n=0}^{\infty}\Big(V_{(n+2,0,\dots,0)} \oplus V_{(n+2,1,0,\dots,0)} \oplus V_{(n+2,2,0,\dots,0)}\Big) \,.
\end{align}
It is evident what the answer is for general spin-\(k\):
\begin{equation}\label{eq:massive_spin-k}
\boxed{\text{Spin-}k:~~\text{Ind}_{SO(d-1)}^{SO(d)}\big(V_{(k,0,\dots,0)}\big) = \bigoplus_{n=0}^{\infty}\bigoplus_{m=0}^k V_{(n+k,m,0,\dots,0)} }
\end{equation}

\subsubsection{Fermions}
Recall the rule for labeling \(SO(d)\) representations:
\begin{equation}
l = (l_1,\dots,l_r)~~\text{with}~~l_i \in \mathbb{Z}/2,~~l_i - l_{i+1}\in \mathbb{Z},~\text{and}~l_1\ge l_2 \ge \cdots \ge l_{r-1} \ge \abs{l_r},
\end{equation}
where \(r = \lfloor d/2 \rfloor\) and the absolute value is only for \(SO(2r)\).

\subsubsection*{Fermions in \(d=4\)}
In \(d=4\) the massive little group is \(SO(3)\) and representations are labeled by a single number \(\bar{l} = (\bar{l}_1)\). The \(SO(4)\) representations are labeled by \(l = (l_1,l_2)\). Upon specifying a \(\bar{l}\) the Frobenius reciprocity conditions read:
\begin{equation}\label{eq:FroRec_d4}
  l_1 \ge \bar{l} \ge \abs{l_2}
\end{equation}
\begin{description}
\item[Dirac fermion $\bar{l}=(\frac{1}{2})$:] The Frobenius reciprocity conditions~\eqref{eq:FroRec_d4} read \( l_1 \ge \frac{1}{2} \ge \abs{l_2}\), which are readily solved to give
\begin{equation}
\text{Ind}_{SO(3)}^{SO(4)}\big(V_{\frac{1}{2}}\big) = \bigoplus_{n=0}^{\infty}\bigg(\frac{1}{2}+n,\frac{1}{2}\bigg) \oplus \bigg(\frac{1}{2} + n,-\frac{1}{2} \bigg). 
\end{equation}
\item[Massive gravitino $\bar{l} = (\frac{3}{2})$:] We solve \(l_1 \ge \frac{3}{2} \ge \abs{l_2}\), giving
\begin{align}
    R_{3/2} = \text{Ind}_{SO(3)}^{SO(4)}\big(\tfrac{3}{2}\big) = \bigoplus_{n=0}^{\infty} \bigg\{ &\underbrace{\bigg[ \bigg(\frac{3}{2}+n,\frac{3}{2}\bigg) \oplus \bigg(\frac{3}{2}+n, -\frac{3}{2}\bigg)\bigg]}_{\text{massless spin 3/2}} \nonumber \\
    &\oplus \underbrace{\bigg[ \bigg(\frac{3}{2}+n,\frac{1}{2}\bigg) \oplus \bigg(\frac{3}{2}+n,-\frac{1}{2}\bigg) \bigg]}_{\text{``soft'' spin 1/2}} \bigg\} \,.
\end{align}
A massive spin \(3/2\) particle has four polarizations; at high energies, these polarizations can be viewed as two coming from a massless spin 3/2 with the other two coming from ``eating'' a massless spin \(1/2\). In other words, this decomposition reflects the Higgs mechanism---as does Eq.~\eqref{eq:massive_spin-k} and the spin-\(k\) examples considered above. We elaborate on this below in subsection~\ref{app:spm_and_higgs}, where we also explain the terminology ``soft''.
\item[General half-integer spin:] Taking \(\bar{l} = \big(k+\frac{1}{2}\big) = \big(\frac{2k+1}{2}\big)\) we find
\begin{equation}
R_{\left(\frac{2k+1}{2}\right)} = \bigoplus_{n=0}^{\infty}\bigoplus_{m=0}^k\bigg[\bigg(\frac{2k+1}{2}+n,\frac{2m+1}{2}\bigg) \oplus \bigg(\frac{2k+1}{2}+n,-\frac{2m+1}{2}\bigg)\bigg] \,.
\end{equation}
\end{description}

\subsection{Massive SPMs and the Higgs mechanism} \label{app:spm_and_higgs}

At high-energies, a massive particle looks massless. In this regime, the polarization states become reducible, up to corrections of \(m/E\). A familiar and useful example of this phenomena is the Goldstone equivalence theorem, which allows us to replace the longitudinal mode of a vector with the Goldstone boson giving it mass.

We therefore expect that a massive SPM can nearly be decomposed into massless SPMs. we say ``nearly'' because the massless SPMs in the ``decomposition'' may not exactly correspond to the usual massless particles due to soft relations. For example, the longitudinal mode of a massive vector can be interpolated by a pion field \(\pd_{\m}\pi\): the pion field \(\pi(x)\) is ``nearly'' the same as a normal massless scalar, except its constant mode is fixed so that the single particle module starts at \(\pd_{\m}\pi\) instead of \(\pi\), \(R_{\pd\pi} = (\pd_{\m}\pi, \pd_{\{\m}\pd_{\n\}}\pi,\dots)\) (\(R_{\pd\pi}\) is what we called \(R_u\), Eq.~\eqref{eqn:pion_SPM}, in the main text). We will call this a ``soft'' scalar, and more generally refer to analogous particles interpolating longitudinal modes as soft particles.

What might we guess for a massive graviton? Likely it will contain a massless spin-2, spin-1, and spin-0. And just like the massive vector, it is reasonable to anticipate that the massless vector and scalar are soft in a certain sense. This intuition is correct; to formalize it, we first need to know the mode decomposition for \textit{massless} particles, which we establish with the following proposition:\\

\noindent \textbf{Proposition 1:} The mode decomposition of the SPM for a massless, spin-\(k\) particle is given by
\begin{equation}\label{eq:massless_spin-k}
  R_{\text{spin-}k,\text{massless}} = \bigoplus_{n=0}^{\infty}V_{(k+n,k,0,\dots,0)} \,.
\end{equation}
\textit{Proof:} A massless spin-\(k\) interpolating field \(A_{\m_1\dots \m_k}\) has a corresponding field strength \(F_{\m_1\dots \m_k \n_1 \dots \n_k}\), transforming in the \((k,k,0,\dots,0)\) representation of \(SO(d)\). The field strength obeys (1) a massless condition, from the Klein-Gordon equation \(\pd^2 F = 0\), (2) a transverse condition \(D^\m F_{\m \m_2\dots \n_k} =0\), and (3) various Bianchi identities, which in practical terms prevent anti-symmetrization of \(D\) on \(F\), \(D_{[\r}F_{\m_1]\m_2\dots\n_k} = 0\). Taken together, these imply that we can only add traceless, symmetric derivatives on top of the field strength. The SPM therefore takes the form \((F_{\m_1\dots \n_k}, D_{\{\r}F_{\m_1\}\dots\n_k}, D_{\{\r_1}D_{\r_2}F_{\m_1\}\dots\n_k}, \dots)\), and~\eqref{eq:massless_spin-k} follows.\\

Comparing the massive and massless SPM decompositions, eqs.~\eqref{eq:massive_spin-k} and~\eqref{eq:massless_spin-k}, we see precisely how the Higgs mechanism shows up. It can be summarized by the following figure:
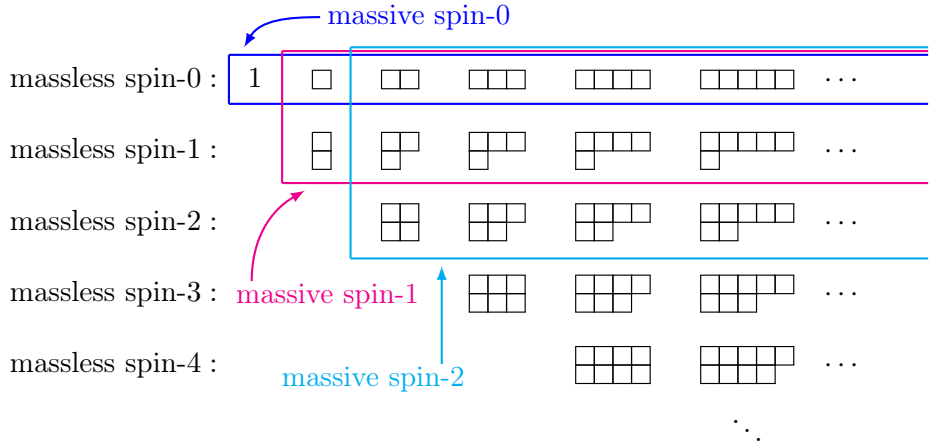
\begin{center}
  \begin{tikzpicture}[>=latex]
    \node (A) at (0,0) {\(\def\arraystretch{1.8} \ytableausetup{boxsize=.6em,centertableaux}
    \begin{array}{rccccccc}
      \text{massless spin-}0: & ~~1~~ & ~~\ydiagram{1}~~ & ~~\ydiagram{2}~~ & ~~\ydiagram{3}~~ & ~~\ydiagram{4}~~ & ~~\ydiagram{5}~~ & \cdots \\
      \text{massless spin-}1: & & \ydiagram{1,1} & \ydiagram{2,1} & \ydiagram{3,1} & \ydiagram{4,1} & \ydiagram{5,1} & \cdots \\
      \text{massless spin-}2: & & & \ydiagram{2,2} & \ydiagram{3,2} & \ydiagram{4,2} & \ydiagram{5,2} & \cdots \\
      \text{massless spin-}3: & & & & \ydiagram{3,3} & \ydiagram{4,3} & \ydiagram{5,3} & \cdots \\
      \text{massless spin-}4: & & & & & \ydiagram{4,4} & \ydiagram{5,4} & \cdots \\
       & & & & & & \ddots &
    \end{array}
    \)};

    \draw[-,blue,thick] (-2.7,2.6) to (6.5,2.6);
    \draw[-,blue,thick] (-2.7,2.6) to (-2.7,1.95);
    \draw[-,blue,thick] (-2.7,1.95) to (6.5,1.95);
    
    \draw[-,magenta,thick] (-2,.9) to (6.5,.9);
    \draw[-,magenta,thick] (-2,.9) to (-2,2.65);
    \draw[-,magenta,thick] (-2,2.65) to (6.5,2.65);
    
    \draw[-,cyan,thick] (-1.1,-.1) to (6.5,-.1);
    \draw[-,cyan,thick] (-1.1,-.1) to (-1.1,2.7);
    \draw[-,cyan,thick] (-1.1,2.7) to (6.5,2.7);

    \draw[<-,blue,thick] (-2.5,2.7) to [out=60,in=180] (-1.5,3.1);
    \node at (-.2,3.1) {\textcolor{blue}{massive spin-0}};

    \draw[<-,magenta,thick] (-1.7,.8) to [out= 210,in=90] (-2.4,-.4);
    \node at (-1.4,-.6) {\textcolor{magenta}{massive spin-1}};

    \draw[<-,cyan,thick] (.1,-.2) to (.1,-1.5);
    \node at (-.8,-1.7) {\textcolor{cyan}{massive spin-2}};
  \end{tikzpicture}
\end{center}
For example, we see that the massive vector \(R_A\) in Eq.~\eqref{eq:massive_spin-1} decomposes into a massless vector plus a soft scalar, while the massive graviton in Eq.~\eqref{eq:massive_spin-2} decomposes into a massless spin-2 plus a soft spin-1 plus a (super-)soft scalar mode (the Galileon mode). To associate a formula to the figure, we need to make up some notation. Let \(R_{A^{(k)}}\) denote the massive spin-\(k\) SPM, let \(R_{F^{(k)}}\) denote the massless spin-\(k\) SPM, and let \(R_{\partial^mF^{(k)}}\) denote the ``\(m\)-soft'' massless spin-\(k\) SPM (i.e. \(R_{F^{(k)}}\) with the first \(m\) modes removed). For example, in this notation the massive graviton SPM decomposes as \(R_{A^{(2)}}   = R_{F^{(2)}} \oplus R_{\pd F^{(1)}} \oplus R_{\pd^2 F^{(0)}}\). In general, a massive spin-\(k\) SPM decomposes as
\begin{equation}\label{eq:MassiveSpinkSPMdecomp}
R_{A^{(k)}} = \bigoplus_{m=0}^k R_{\partial^mF^{(k-m)}} \,.
\end{equation}

\subsection{Characters}
Important to the Hilbert series analysis is computation of the character \(\ch_{\Ph}\) corresponding to the single particle module \(R_{\Ph}\),
\begin{equation}
\ch^{}_{\Ph}(q,x) = \sum_{l\in \L_{\Ph}} q^{\D(l)}\ch_l^{}(x) \,,
\end{equation}
where \(\ch_l^{}(x)\) is the \(SO(d)\) character of \(V_l\) appearing in the decomposition \(R_{\Ph} = \bigoplus_{l \in \L_{\Ph}} V_l\), and \(\D(l)\) is the scaling dimension of \(V_l\). If the scaling dimension of the field \(\Ph\) is \(\D_{\Ph}\), then the scaling dimension \(\D(l)\) is simply \(\D_{\Ph}\) plus the number of derivatives added to reach the representation \(V_l\).\footnote{While it's often obvious what \(\D(l)\) is in any specific example, it is possible to give a general and explicit formula for \(\D(l)\). Because all particles obey the Klein-Gordon equation \(-\pd^2\Ph = m^2 \Ph\) (a mass-shell condition) and all polarizations are transverse \(\pd^{\m}\Ph_{\m\a\b\dots} = 0\), it implies that no derivatives are contracted in the SPM. Thus, adding a derivative will always change the \(SO(d)\) representation in some way; said another way, it will always add another box to the Young diagram. Therefore the number of boxes in the Young diagram \(\abs{l} \equiv l_1+l_2 + \cdots + \abs{l_r}\) (\(\abs{l}\) is called the length of the partition \(l\)) essentially counts the number of derivatives. Specifically, if we let \(l_{\Ph}\) denote the partition corresponding to the first mode in the SPM, then the scaling dimension of \(V_l \subset R_{\Ph}\) is given by \(\D(l) = \D_{\Ph} + \abs{l} - \abs{l_{\Ph}}\).} The \(SO(d)\) character \(\ch_l^{}(x)\) can be obtained from the Weyl character formula, \textit{e.g.}~\cite{FultonHarris} (this is also reviewed in App. A of~\cite{Henning:2017fpj}).

Since the SPMs decompose into infinite sums of \(SO(d)\) modules, the characters \(\ch_{\Ph}^{}\) involve infinite sums of \(SO(d)\) characters (graded by their scaling dimension). Using the Weyl character formula, together with the geometric series expression \(\sum_{k=0}^{\infty} x^k = 1/(1-x)\), these sums can frequently be performed directly, especially with the help of a program such as \texttt{Mathematica}.

As examples, let us compute the characters for massive spin-\(k\) particles. For a massive vector \(A_{\m}\) we have
\begin{equation}
  R_A = \bigoplus_{n=0}^{\infty} \Big(V_{(n+1,1,0,\dots,0)}\oplus V_{(n+1,0,\dots,0)}\Big) \simeq \begin{pmatrix} F \\ \pd F \\ \pd^2F \\ \vdots \end{pmatrix} \oplus \begin{pmatrix} \pd \ph \\ \pd^2 \ph \\ \pd^3\ph \\ \vdots \end{pmatrix} = R_F \oplus R_{\pd\ph} \,.
\end{equation}
Assigning scaling dimension \([A_{\m}] = \D_A\), the character is given by
\begin{align}\ytableausetup{boxsize=.35em,centertableaux}\label{eq:char_massive_spin-1} 
\ch_A(q,x) &= \sum_{n=0}^{\infty}q^{\D_A+n}\Big(\ch_{(n+1,0,\dots,0)}(x) + q\, \ch_{(n+1,1,0,\dots,0)}(x)\Big) \nonumber \\[5pt]
&= q^{\D_A}(1-q^2)\big(\ch_{(1,0,\dots,0)}(x) - q \big)P(q,x) \,,
\end{align}
which reproduces Eq.~\eqref{eqn:A_char} of the main text. For a massive graviton we have (assigning scaling dimension \([g_{\mu\nu}] = \D_g\))
\begin{align}
\ch_g(q,x) &= \sum_{n=0}^{\infty}q^{\D_g+n}\Big(\ch^{}_{(n+2,0,\dots,0)}(x) + q\, \ch^{}_{(n+2,1,0,\dots,0)}(x)+ q^2 \, \ch^{}_{(n+2,2,0,\dots,0)}(x)\Big) \nonumber \\[5pt]
&= q^{\D_g}(1-q^2)\big(\ch^{}_{(2,0,\dots,0)}(x) - q\, \ch^{}_{(1,0,\dots,0)} \big)P(q,x) \,. \label{eq:char_massive_spin-2}
\end{align}
Comparing to the schematic picture in Eq.~\eqref{eqn:vector_SPM_schematic}, we see a very similar picture reflected in~\eqref{eq:char_massive_spin-2}. In fact, this pattern holds for arbitrary massive spin-\(k\), where we find:
\begin{align}
\ch_{A^{(k)}}^{}(q,x) &= \sum_{n=0}^{\infty}\sum_{m=0}^kq^{\D_0+n+m}\ch^{}_{(n+k,m,0,\dots,0)}(x) \nonumber \\[5pt]
&= q^{\D_0}(1-q^2)\big(\ch^{}_{(k,0,\dots,0)}(x) - q\, \ch^{}_{(k-1,0,\dots,0)} \big)P(q,x) \,. \label{eq:char_massive_spin-k}
\end{align}
This equality makes sense because it reflects that the spin-\(k\) interpolating field \(A^{(k)} = A^{(k)}_{\m_1\dots \m_k}\) (\(A^{(1)}_{\m} \equiv A_{\m}\) for a vector, \(A^{(2)}_{\m\n} \equiv g_{\m\n}\) for a graviton) obeys the following two equations of motion: (1) a mass condition \((\pd^2 + m^2)A^{(k)} = 0\) and (2) a transverse condition \(\pd^{\m}A_{\m\m_2 \dots \m_k} = 0\). This is the spin-\(k\) generalization of equations~\eqref{eqn:proca},~\eqref{eqn:vector_SPM_schematic}, and~\eqref{eqn:A_char} in the main text.

Interestingly, the characters for massive spin-\(k\) particles~\eqref{eq:char_massive_spin-k} are easier to obtain than their massless counterparts. At the heart of this is the fact that the equations of motion for a massive particle are in a sense ``simpler'', consisting of the Klein-Gordon equation together with the transverse condition, while massless particles in addition have to obey Bianchi identities (see~\cite{Bekaert:2006py} for a treatment of Poincar\'e representations in \(d\) spacetime dimensions). In particular, despite knowing the mode decomposition of massless SPMs (e.g. Eq.~\eqref {eq:massless_spin-k}), we have been unable to find a general formula for massless spin-\(k\) characters in arbitrary dimension \(d\), although we stress that specific cases are easily found using the Weyl character formula and \texttt{Mathematica}.

Above we showed how the mode decomposition of the massive SPM reflects the Higgs mechanism, i.e. how the massive spin-\(k\) SPM decomposes into a massless spin-\(k\) SPM plus a sequence of soft massless spin-\(j\) SPMs, \(j = 0, 1, \dots, k-1\). The characters \(\ch_{A^{(k)}}(q,x)\) of course reflect this decomposition, giving analogous expressions to Eq.~\eqref{eqn:HiggsMechanism}. Similar to the study of HEFT in the main text, it is frequently convenient to explicitly separate out these longitudinal components by assigning them independent gradings. Unfortunately, we cannot provide a general formula for arbitrary spin-\(k\) in general dimension \(d\), since we have not been able to find a general formula for the massless characters; however, in any specific example it is straightforward and relatively simple to find the analog of Eq.~\eqref{eqn:HiggsMechanism} for the case at hand.

Finally, one can also account for the effects of discrete spacetime symmetries such as parity and charge conjugation following the procedures outlined in~\cite{Henning:2017fpj} (see Appendix C) and~\cite{Graf:2020yxt}.

\bibliographystyle{JHEP}
\bibliography{references}

\end{document}

%% file: preamble.tex
\def\a{\alpha}
\def\b{\beta}

\def\D{\Delta}

\def\e{\epsilon}

\def\L{\Lambda}
\def\m{\mu}
\def\n{\nu}
\def\s{\sigma}
\def\r{\rho}

\def\ph{\phi}
\def\Ph{\Phi}

\def\ch{\chi}
\def\abs#1{\left| #1 \right|}
\def\pd{\partial}

\newcommand{\be}{\begin{eqnarray}}
\newcommand{\ee}{\end{eqnarray}}

\makeatletter
\newsavebox\myboxA
\newsavebox\myboxB
\newlength\mylenA
\newcommand*\xoverline[2][0.75]{%
    \sbox{\myboxA}{$\m@th#2$}%
    \setbox\myboxB\null
    \ht\myboxB=\ht\myboxA%
    \dp\myboxB=\dp\myboxA%
    \wd\myboxB=#1\wd\myboxA
    \sbox\myboxB{$\m@th\overline{\copy\myboxB}$}
    \setlength\mylenA{\the\wd\myboxA}
    \addtolength\mylenA{-\the\wd\myboxB}%
    \ifdim\wd\myboxB<\wd\myboxA%
       \rlap{\hskip 0.5\mylenA\usebox\myboxB}{\usebox\myboxA}%
    \else
        \hskip -0.5\mylenA\rlap{\usebox\myboxA}{\hskip 0.5\mylenA\usebox\myboxB}%
    \fi}
\makeatother

\tikzset{SU/.style={draw,circle}}
\tikzset{block/.style = {draw, text centered, 
              minimum height=0.8cm, inner sep=0}}
\tikzset{ball/.style = {circle, draw, text centered, anchor=north, inner sep=0}}
\tikzset{ballc/.style = {circle, draw, text centered, inner sep=0}}
\tikzset{->-/.style={decoration={
  markings,
  mark=at position #1 with {\arrow{>}}},postaction={decorate}}}

\tikzset{->>-/.style={decoration={
  markings,
  mark=at position #1 with {\arrow{>>}}},postaction={decorate}}}